\documentclass[aip,reprint]{revtex4-1}

\usepackage{caption}

\usepackage{graphicx}
\usepackage{dcolumn}
\usepackage{bm}

\usepackage[utf8]{inputenc}
\usepackage[T1]{fontenc}
\usepackage{mathptmx}
\usepackage{etoolbox}

\usepackage{caption}
\usepackage{lineno}
\usepackage[section]{placeins} 
\usepackage{float} 
\usepackage{subcaption} 
\usepackage{gensymb}

\usepackage{mhchem} 

\usepackage{hyperref}
\hypersetup{colorlinks,allcolors=black}

\newcommand\leff{l_{\rm{eff}}}

\usepackage{pifont}
\usepackage{color}

\makeatletter
\def\@email#1#2{%
 \endgroup
 \patchcmd{\titleblock@produce}
  {\frontmatter@RRAPformat}
  {\frontmatter@RRAPformat{\produce@RRAP{*#1\href{mailto:#2}{#2}}}\frontmatter@RRAPformat}
  {}{}
}%
\makeatother

\begin{document}


\title[Calorimetric Wire Detector for Measurement of Atomic Hydrogen Beams]{Calorimetric Wire Detector for Measurement of Atomic Hydrogen Beams}
\newcommand{\Arlington}{\affiliation{Department of Physics, University of Texas at Arlington, Arlington, TX 76019, USA}}
\newcommand{\Mainz}{\affiliation{Institute for Physics, Johannes Gutenberg University Mainz, 55128 Mainz, Germany}}
\newcommand{\MIT}{\affiliation{Laboratory for Nuclear Science, Massachusetts Institute of Technology, Cambridge, MA 02139, USA}}
\newcommand{\PennStateARL}{\affiliation{Applied Research Laboratory, Pennsylvania State University, University Park, PA 16802, USA}}
\newcommand{\PennState}{\affiliation{Department of Physics, Pennsylvania State University, University Park, PA 16802, USA}}
\newcommand{\PNNL}{\affiliation{Pacific Northwest National Laboratory, Richland, WA 99354, USA}}
\newcommand{\Yale}{\affiliation{Wright Laboratory and Department of Physics, Yale University, New Haven, CT 06520, USA}}
\newcommand{\Livermore}{\affiliation{Lawrence Livermore National Laboratory, Livermore, CA 94550, USA}}
\newcommand{\Case}{\affiliation{Department of Physics, Case Western Reserve University, Cleveland, OH 44106, USA}}
\newcommand{\Heidelberg}{\affiliation{Institute for Theoretical Astrophysics, Heidelberg University, 69120 Heidelberg, Germany}}
\newcommand{\Illinois}{\affiliation{Department of Physics, University of Illinois Urbana-Champaign, Urbana, IL 61801, USA}}
\newcommand{\Indiana}{\affiliation{Center for Exploration of Energy and Matter and Department of Physics, Indiana University, Bloomington, IN, 47405, USA}}
\newcommand{\KIT}{\affiliation{Institute of Astroparticle Physics, Karlsruhe Institute of Technology, 76021 Karlsruhe, Germany}}
\newcommand{\Pitt}{\affiliation{Department of Physics \& Astronomy, University of Pittsburgh, Pittsburgh, PA 15260, USA}}
\newcommand{\Washington}{\affiliation{Center for Experimental Nuclear Physics and Astrophysics and Department of Physics, University of Washington, Seattle, WA 98195, USA}}
\newcommand{\Ghent}{\affiliation{Department of Physics and Astronomy, Ghent University, 9000 Gent, Belgium}}

\author{M.~Astaschov}\Mainz
\author{S.~Bhagvati}\PennState
\author{S.~Böser}\Mainz
\author{M.~J.~Brandsema}\PennStateARL
\author{R.~Cabral}\Indiana
\author{C.~Claessens}\Washington
\author{L.~de~Viveiros}\PennState
\author{S.~Enomoto}\Washington
\author{D.~Fenner}\Mainz
\author{M.~Fertl}\Mainz
\author{J.~A.~Formaggio}\MIT
\author{B.~T.~Foust}\PNNL
\author{J.~K.~Gaison}\PNNL
\author{P.~Harmston}\Illinois
\author{K.~M.~Heeger}\Yale
\author{M.~B.~Hüneborn}\Mainz
\author{X.~Huyan}\PNNL
\author{A.~M.~Jones}\PNNL
\author{B.~J.~P.~Jones}\Arlington
\author{E.~Karim}\Pitt
\author{K.~Kazkaz}\Livermore
\author{P.~Kern}\Mainz
\author{M.~Li}\MIT
\author{A.~Lindman}\Mainz
\author{C.-Y.~Liu}\Illinois
\author{A.~Marsteller}\Washington
\author{C.~Matthé}\Mainz
\author{R.~Mohiuddin}\Case
\author{B.~Monreal}\Case
\author{B.~Mucogllava}\Mainz
\author{R.~Mueller}\PennState
\author{A.~Negi}\Arlington
\author{J.~A.~Nikkel}\Yale
\author{N.~S.~Oblath}\PNNL
\author{M.~Oueslati}\Indiana
\author{J.~I.~Peña}\MIT
\author{W.~Pettus}\Indiana
\author{R.~Reimann}\Mainz
\author{A.~L.~Reine}\Indiana
\author{R.~G.~H.~Robertson}\Washington
\author{D.~Rosa~De~Jesús}\PNNL
\author{L.~Saldaña}\Yale
\author{P.~L.~Slocum}\Yale
\author{F.~Spanier}\Heidelberg
\author{J.~Stachurska}\MIT\Ghent
\author{Y.-H.~Sun}\Case
\author{P.~T.~Surukuchi}\Pitt
\author{A.~B.~Telles}\Yale
\author{F.~Thomas}\Mainz
\author{L.~A.~Thorne}\Mainz
\author{T.~Thümmler}\KIT
\author{W.~Van~De~Pontseele}\MIT
\author{B.~A.~VanDevender}\Washington\PNNL
\author{T.~E.~Weiss}\Yale
\author{M.~Wynne}\Washington
\author{A.~Ziegler}\PennState
\collaboration{{Project 8 Collaboration}\noaffiliation
}%
 \email{chmatthe@uni-mainz.de}


\date{\today}

\begin{abstract}

A calorimetric detector for minimally disruptive measurements of atomic hydrogen beams is described. The calorimeter measures heat released by the recombination of hydrogen atoms into molecules on a thin wire. As a demonstration, the angular distribution of a beam with a peak intensity of $\approx 10^{16} \,{\rm{atoms}}/{(\rm{cm}^2 \rm{s})}$ is measured by translating the wire across the beam. The data agree well with an analytic model of the beam from the thermal hydrogen atom source.
Using the beam shape model, the relative intensity of the beam can be determined to 5\% precision or better at any angle. 
\end{abstract}

\maketitle

\section{Introduction}

The work presented in this paper was carried out in the context of development of an atomic tritium source for the Project 8 experiment,\cite{Esfahani2017,Project8_collab_2022} which aims to employ atomic tritium in combination with the novel technology of Cyclotron Radiation Emission Spectroscopy\cite{Monreal2009} to achieve a neutrino mass sensitivity of $m_{\beta} < 40\,{\rm meV/\textit{c}^2}$. As an intermediate goal an atomic hydrogen source capable of producing on the order of $10^{19} \, \rm{atoms}/\rm{s}$ is being developed, which will eventually be adapted to become the atomic tritium source.

As part of this effort, detectors capable of monitoring the resulting atomic hydrogen beam will be required. Traditionally this could be achieved using a high resolution quadrupole mass spectrometer (QMS). However, situations are anticipated in which a QMS would not be ideal, in particular in the presence of high magnetic fields and tight space constraints.
To complement measurements with a QMS in such cases, we are developing a calorimetric wire detector, which measures the heat of atoms recombining into molecules.

The concept of using such detectors for atomic hydrogen beams has previously been explored \cite{ Trofimov2005,Ugur2012}. A thin wire is strung across the beam, such that it intercepts a small slice of the beam. A sizable fraction \cite{Winkler1998} of the atoms that hit the wire stick to its surface, where they can recombine into molecules, releasing 4.46 eV per molecule \cite{Cheng2018}. A portion of this energy is absorbed by the wire causing a slight heating, which can be detected by measuring the corresponding increase in resistance.

An advantage of the wire calorimeter method is that, by using very thin wires, the detection of the beam can be accomplished without meaningfully disturbing the beam.
This is useful for online monitoring a beam that may continue to be used for its primary purpose downstream of the detector.
The wire detector can also be used at higher pressures than are required for mass spectrometry. This is desirable when scaling to more intense hydrogen beams without a commensurate increase in pumping on the vacuum system.
The  Project 8 experiment will eventually employ strong magnetic fields\cite{Project8_collab_2022},which will cause large issues for the sensitivity of a QMS\cite{Syed2013} due to the free electrons and ions involved in its operation. It is expected that the wire detector should operate largely unaffected by magnetic fields, though this has not yet been shown experimentally.  Finally, wire detectors can be built on a very thin supporting structure, such that they may be  employed in confined spaces, where fitting a QMS might be challenging.

\begin{figure}[htbp]
            \includegraphics[width=1\linewidth]{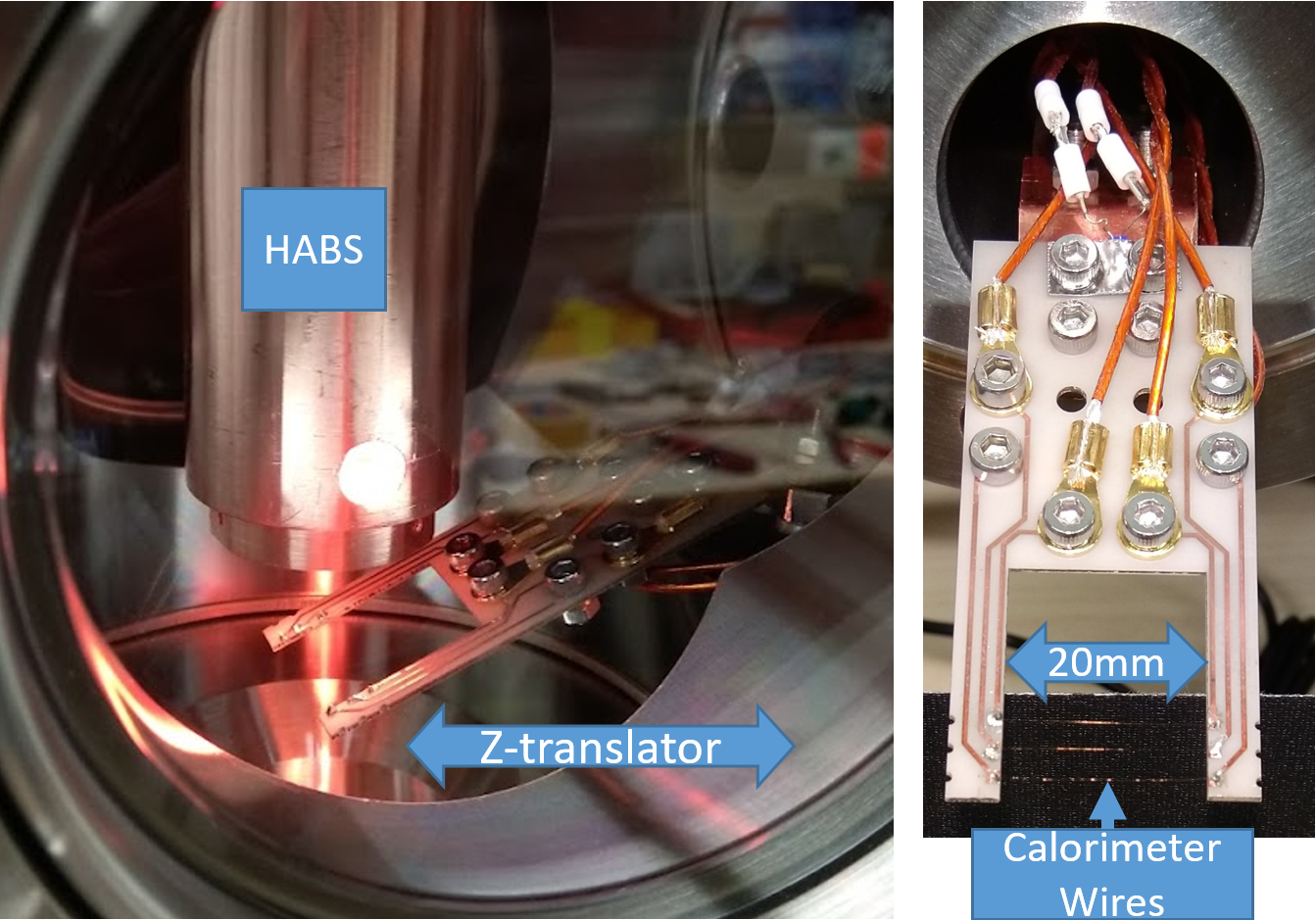}
            \caption{\label{fig:wire_detector}On the left is a view into the vacuum chamber containing both the hydrogen atomic beam source (HABS) and the wire detector. The HABS can be seen to be glowing as it does when operated at high temperatures. The z-axis, along which the wire detector can be moved, is indicated. A frontal view of the wire detector is shown on the right. Three wires are attached to the detector board for redundancy. The rearmost wire is used for all the data presented in this paper.}
\end{figure}

The calorimetric wire detector used in this study, depicted in Figure \ref{fig:wire_detector}, consists of a $5\,$µm  thick gold-coated tungsten wire that is strung across a $20\, \rm{mm}$ wide gap. 
On each side of the gap, the wire is mounted on the aluminum nitride PCB by soldering it to copper traces leading to the readout electronics.
The thickness of the wire was chosen to be the thinnest that could be readily procured and assembled into the detector. A thinner wire has higher base resistance as well as a lower heat conductivity such that, for a given heating power, temperature changes to the wire and resulting resistance changes are maximized. For the same reason a longer wire to further increase the aspect ratio of the detector is preferable. The length of the wire used here was chosen to fit within the space available in the vacuum chamber and to pass through a CF40 flange. 

The wire resistance is measured by applying a constant current to the wire and recording the voltage drop over the wire with a digital multimeter. The wires typically have a room temperature resistance of $\approx 65 \, \Omega$ rising to as much as $100 \, \Omega$ during operation.

The detector PCB is mounted on a z-translator stage \footnote{McAllister Bellows-sealed Linear Translator (CF 40)}, such that the wire can be scanned across the hydrogen beam. About 31 mm of travel is available within the confines of the vacuum chamber. This travel covers roughly -15° in one direction and +30° of the beam in the other direction of the capillary axis, as seen from the wire's location. Approximately 25\% of particles emitted from the capillary are in this region, and the beam intensity drops to $\approx 40$\% of its maximum intensity at 20° off-center at the settings presented later in this paper. Due to an aperture, the beam is cut off beyond 26° allowing for measurements with the wire fully outside of the beam.

We have successfully used the wire detector to measure the relative intensity profile of an atomic hydrogen beam containing of order $10^{16} \,{\rm{atoms}}/{(\rm{cm}^2 \rm{s})}$, using a number of data-driven corrections to extract a small heating signal in a complex thermal environment. We have compared the measured profile with theoretical models of the source and found them to be in agreement.

The paper is divided as follows:
we begin with a description of the hydrogen beam we use (Section \ref{sec:Beam source}), followed by the calibration of the wire detector (Section \ref{sec:Calibration}). Then an explanation of various effects influencing the wire detector's resistance will be discussed (Section \ref{sec:Signal components}) and the method for extracting the atomic hydrogen signal from the raw data will be described (Section \ref{sec:signal extraction}). Finally, the reconstruction of the relative beam intensity distribution based on data will be demonstrated (Section \ref{sec:Application to Data}). Multiple appendices are attached to provide more detail on selected specifics.

\section{Hydrogen Beam}
\label{sec:Beam source}
\subsection{Source}
\label{sec:Beam source sub}

For development work, including the results presented here, we use a commercially available "Hydrogen Atomic Beam Source" (HABS)\footnote{https://www.mbe-komponenten.de/products/mbe-components/gas-sources/habs.php}. This is a thermal hydrogen dissociator manufactured by \textit{MBE Komponenten GmbH} that produces atomic hydrogen by passing a flow of molecular hydrogen through a tungsten capillary that is heated to around 2200 K \cite{Tschersich2008}. The hydrogen flow through the source is controlled by a mass flow controller\footnote{Alicat Model: MCE-20SCCM-D-DB15K } in the range of $0.002-20$ standard cubic centimeter per minute (sccm). Before entering the source, the gas flows through a purifier\footnote{Entegris GateKeeper MC1-904F} which removes any water, oxygen or hydrocarbon contaminants.

The HABS capillary is located 35 mm above the wire detector. 
At the settiKazngs used in the work presented here, 1 sccm of $\rm{H_2}$ flow at 2200 K, the source produces a beam intensity of order $10^{16} \,{\rm{atoms}}/{\rm{cm}^2\rm{s}}$.\cite{Tschersich2008}  In the ideal case where all atoms landing on the wire recombine, releasing 4.46 eV per molecule, this corresponds to a heat load of about $4 \,\rm{mW}/\rm{cm}^2$ or total of 4 $\mu$W along the length of the wire.

Hydrogen is emitted from the HABS as a diverging beam, with an initially undetermined fraction of dissociation~\cite{Tschersich2000}. 
Since the wire is not a point detector, it always integrates a portion of the beam along its length. The change in beam intensity along this length must be  modeled.

We derive the HABS beam output intensity distribution  $j_{\rm{HABS}}$ based on a theoretical model\cite{Tschersich1998} modified by the setup geometry. As it will be defined, $j_{\rm{HABS}}$ is the probability density per unit angle $[1/\rm{sr}]$ of finding a particle which leaves the HABS capillary at a certain angle from the capillary axis downstream of the copper cooling shroud covering the HABS.

\begin{figure}[H]
    \centering
    \includegraphics[width=0.9\linewidth]{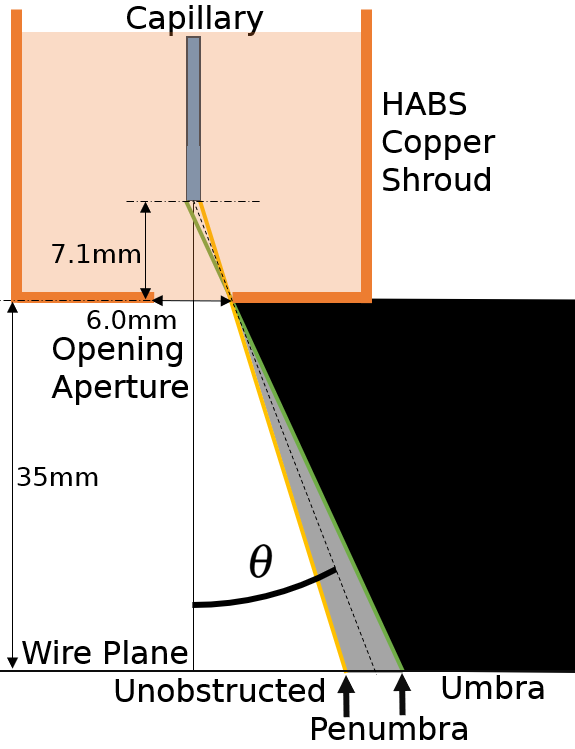}
    \caption{A sketch of the geometry of the HABS copper shroud, which partially and then fully obscures the capillary from which the hydrogen beam originates. The inside edge of the penumbra is located at 19.4° and the outside edge at 26.2°. Beyond the outside edge no portion of the beam is expected to be visible (umbra). The angle $\theta$ is always measured between the capillary axis and a line drawn from the center of the capillary at its front face. Additional clarification on the coordinate definitions is available in Appendix  \ref{sec:Axes Definition}. 
    }
    \label{fig:penumbra_sketch}
\end{figure}

Figure \ref{fig:penumbra_sketch} illustrates the geometry of the HABS. The capillary forms an initial beam which is then modified, when parts of it are obscured by the copper shroud. We call the fraction of the capillary which is visible under a viewing angle $\theta$ relative to the capillary axis $g_{\rm{visible}}(\theta)$. This functions as a geometric correction factor to the initial output intensity distribution of the capillary $j_{\rm{norm}} (\theta ;\leff)$ such that
\begin{equation}
\label{eq:H_profile}
j_{\rm{HABS}}(\theta ; \leff) = j_{\rm{norm}} (\theta ;\leff) \cdot g_{\rm{visible}}(\theta) 
.
\end{equation}
This does include an implicit assumption that gas exiting the capillary has a uniform density across the exit plane. This may not be exactly true, and Monte Carlo simulations indicate this may change the effective $g$ factor by a few percent, depending on the actual gas density distribution. Absent a fully-fledged model for the gas density distribution in the exit plane of the capillary, for the purposes of this paper, we will assume it to be a uniform distribution, such that every point in the capillary exit plane emits gas with equal intensity.

As described in the following text, we construct $j_{\rm{norm}} (\theta ;\leff)$ such that it is a normalized probability density function. Normalization is useful, since the total flow of molecular hydrogen through the source capillary is known and controlled by a mass flow controller.

We start from the theoretical model of beam intensity from a cylindrical capillary, which is adopted from literature and denoted with $j$. Refer to Tschersich et. al. \cite{Tschersich1998} for a comprehensive explanation.  $j$ is a function with a single dimensionless shape parameter called the effective length, $\leff$, which describes the effective ratio of length to diameter of the capillary. $\leff$ differs from the physical aspect ratio of the capillary if gas flow through the capillary is large enough, such that the flow cannot be approximated as transparent molecular flow along the entire length. All gas flows used in our study are large enough that $\leff$ is much shorter than the physical length of the capillary.

\begin{equation}
\begin{array}{rl}
j(\theta) &= j_d(\theta) + j_w(\theta) \\
\text{with} \quad j_d(\theta) &= \cos(\theta) \cdot U(\beta) \\
\text{and} \quad j_w(\theta) &= \frac{4}{3\pi} \left(1 - \frac{1}{2\leff +1} \right) \frac{1}{\leff} \frac{\cos(\theta)^2}{\sin(\theta)} (1 - V(\beta)) \\
&\quad +  \left(1 - \frac{1}{2\leff +1} \right) \cos(\theta) (1- U(\beta)) , 
\end{array}
\end{equation}

\begin{flalign}
\rm{where} & 
\left \{ \begin{array}{ll} \left. \begin{array}{l} U(\beta) = ( 2 \beta - \mbox{sin}(\beta) ) / \pi \\ V(\beta) = \mbox{sin}(\beta)^3 \end{array} \right\} & \theta < \mbox{arctan} (1/\leff) \\ \begin{array}{l} U(\beta) = V(\beta) = 0 \end{array} & \mbox{otherwise} \end{array} \right. \\
\text{and} & \quad  \beta(\theta) = \arccos(\leff \cdot \tan(\theta)) .
\end{flalign}

\begin{figure*}[htbp]
    \centering
    \begin{subfigure}[t]{0.5\textwidth}
        \centering
        \includegraphics[width=1\linewidth]{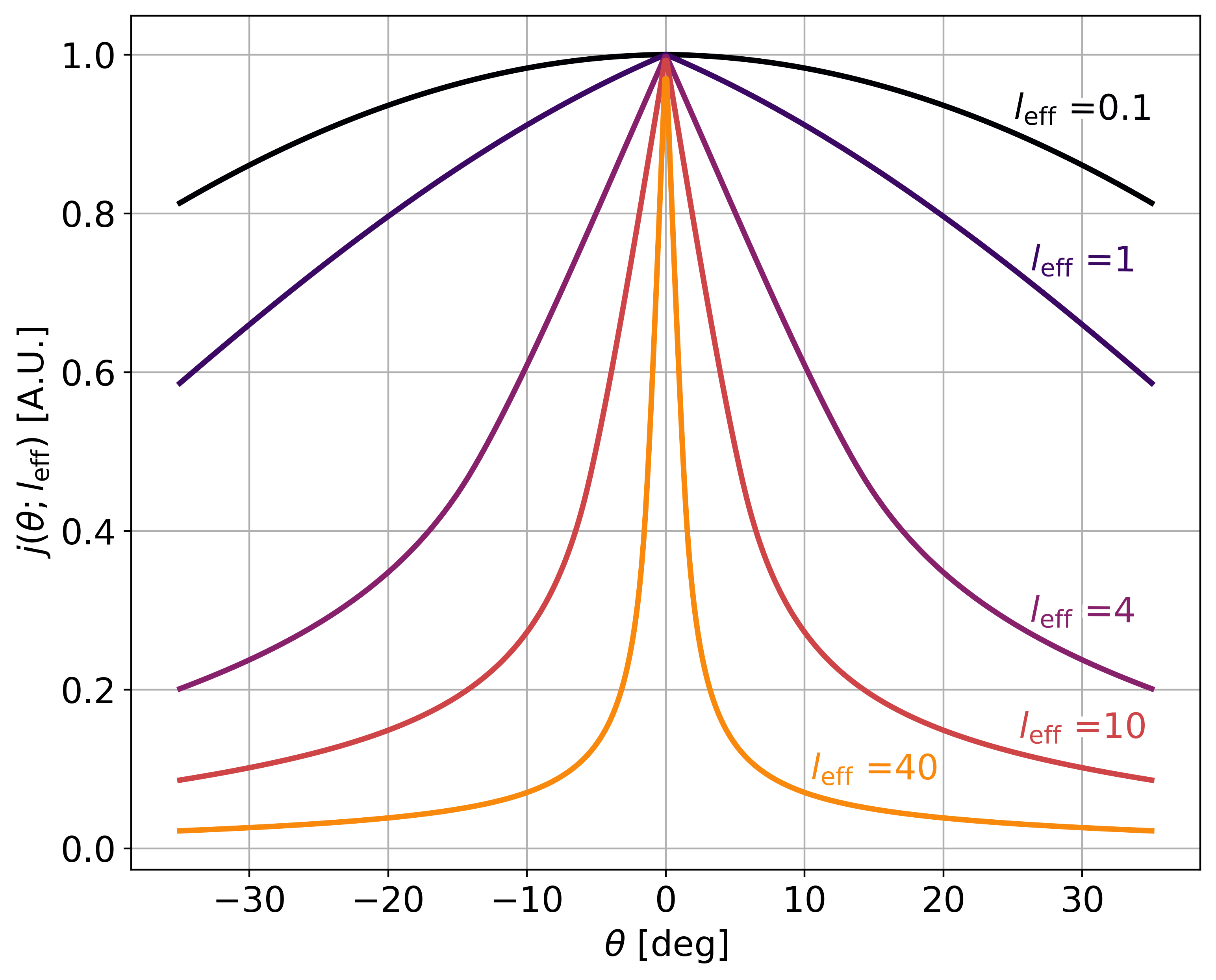}
        \caption{The relative output intensity $j$ as a function of angle from the capillary axis.  $j$ converges to the Lambert cosine law for small $\leff$, equivalent to a zero length aperture. For large $\leff$ it becomes a more centrally focused beam. }
    \end{subfigure}%
    ~ 
    \begin{subfigure}[t]{0.5\textwidth}
        \centering
        \includegraphics[width=1\linewidth]{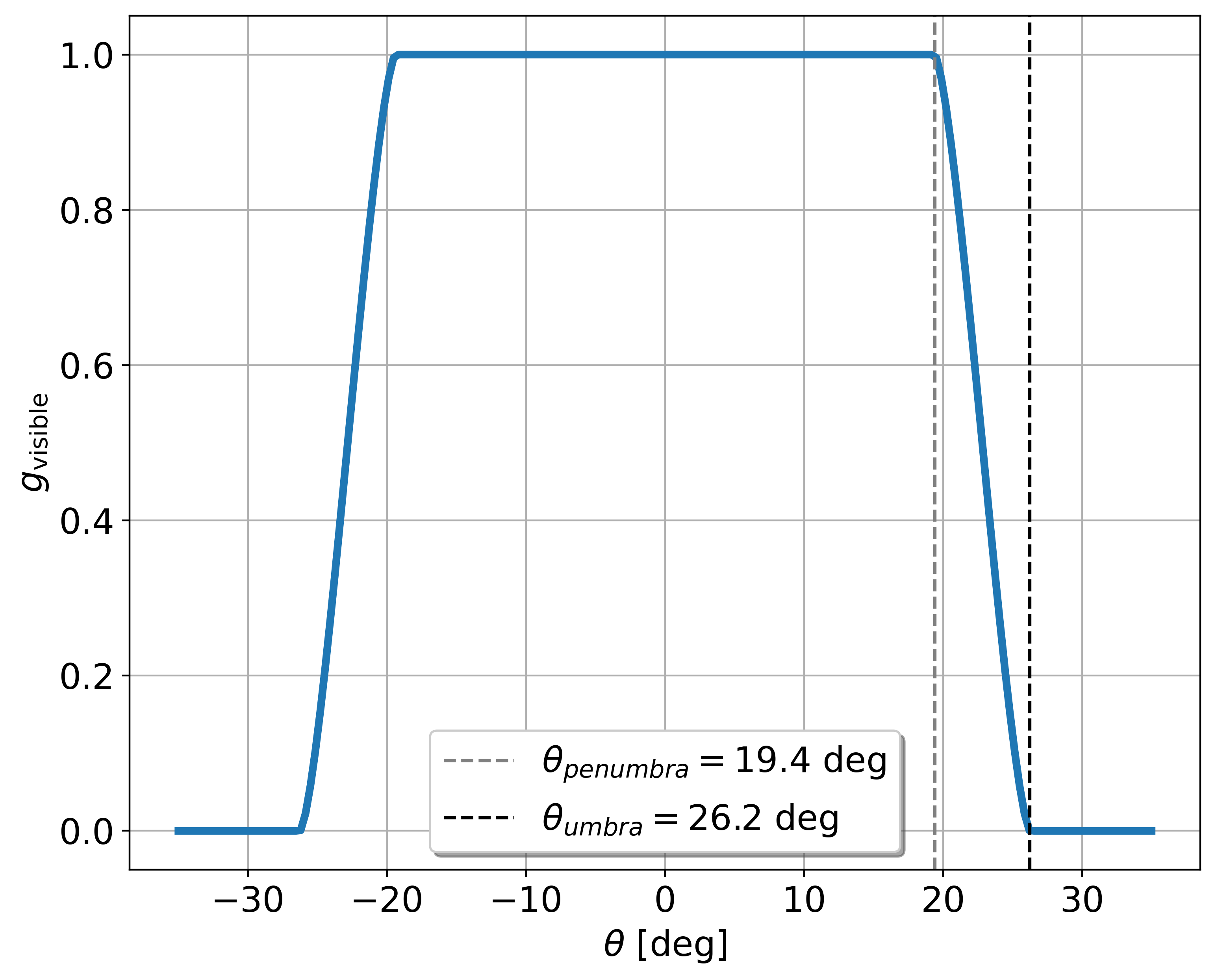}
        \caption{The fraction of the capillary opening that is visible to an observer (in this case the  wire) from the angle $\theta$. At high angles the copper shroud surrounding the HABS capillary obscures first part and then the entire opening, leading to reduced hydrogen flux at those angles.}
    \end{subfigure}
    \caption{The two beam shaping components of $j_{\rm{HABS}}(\theta ; \leff)$ are plotted as a function of angle. On the left the shape of the beam as it exits the source capillary, and on the right the multiplicative geometric modification function $g_{\rm{visible}}$ that is caused by  the exit aperture of the HABS shroud. The final combined beam shape, for parameters matching measurements presented later is shown in Figure \ref{fig:H_profile_penumbra}.
    }
    \label{fig:H_profile}
\end{figure*}

Figure \ref{fig:H_profile} shows how $j$ and $g_{\rm{visible}}$ evolve with angle. $j$ falls off to reduced intensities as the angle increases, decreasing more sharply for larger $\leff$. $g_{\rm{visible}}$ acts as a window function with a smooth transition from fully visible at small angles, to partially shadowed, to fully obstructed at larger angles.


We normalize the intensity profile $j$ by the integral over the emittance hemisphere

\begin{equation}
    j_{\rm{norm}} (\theta, \leff) = \frac{j(\theta, \leff)}{
    \int_0^{\frac{\pi}{2}} \int_0^{2\pi} 
    j(\theta, \leff) \sin{\theta}d\theta d\varphi },
\end{equation}

 It should be noted here that since $j$ is also a function of the effective length $l_{\rm{eff}}$ the normalization integral has to be recalculated for every $l_{\rm{eff}}$. The normalization as described here results in the integral of $j_{\rm{norm}}$ over the hemisphere being equal to one.

 \begin{equation}
    \int_0^{\frac{\pi}{2}} \int_0^{2\pi} 
    j_{\rm{norm}}(\theta, \leff) \sin{\theta}d\theta d\varphi  = 1.
\end{equation}

Normalized this way, $j_{\rm{norm}}$ can be treated as a probability density function for the emission of any single gas particle.

After multiplication with $g_{\rm{visible}}(\theta)$, $j_{\rm{HABS}}(\theta ; \leff)$ is of course no longer normalized to 1. This is the intended behavior, as we can measure how many particles flow through the capillary, but we cannot directly measure how many make it out of the copper shroud.

\subsection{Modeling expected power}
\label{sec:Modeling expected power}

This section will show that a given model for the beam shape such as $j_{\rm{HABS}}$ from Eq.\eqref{eq:H_profile} could be integrated to predict the absolute expected heating power received by the wire. However, due to some poorly-known parameters we only carry out part of the integration and leave the overall heat scaling as a free parameter.

The recombination intensity $I_\mathrm{rec}$ is the power density per unit solid angle $[\rm{W}/\rm{sr}]$ from recombination heating on a surface exposed to an atomic hydrogen beam. It can be expressed as
\begin{equation}  
    I_\mathrm{rec} = \Phi_\mathrm{mol} \cdot 2\alpha_{\rm{dissoc}} \cdot j_{\rm{HABS}}(\theta ; \leff)
    \cdot \gamma_\mathrm{rec}\cdot \beta_\mathrm{rec}\cdot \frac{E_\mathrm{rec}}{2}
    ,
\end{equation}
where $\Phi_\mathrm{mol}$ is the flow of molecules per second into the source, and $\alpha_{\rm{dissoc}}$ is the fraction of these which are dissociated into atoms in the source. $\gamma_\mathrm{rec}$ is the likelihood for an atom that hits the wire to recombine on it, $ \beta_\mathrm{rec}$ is the fraction of the recombination energy that is transferred to the wire if a molecule is produced, and $E_\mathrm{rec}$ is the total energy released when two hydrogen atoms recombine into a molecule.

The values of $\alpha_{\rm{dissoc}}$, $\gamma_\mathrm{rec}$, and $ \beta_\mathrm{rec}$ are all initially unknown. They are difficult to disentangle with currently available data. Some values for the recombination parameters are available from literature\cite{Melin1970}, but with large uncertainties, and it is unclear if they are directly applicable.  Due to the degeneracy caused by including all three as separate parameters, for the purposes of a fit model, all such purely multiplicative parameters will be combined into a single scaling parameter $A =  \Phi_\mathrm{mol} \cdot 2\alpha_{\rm{dissoc}} \cdot \gamma_\mathrm{rec}\cdot \beta_\mathrm{rec}\cdot \frac{E_\mathrm{rec}}{2}$. $A$ is then implicitly a complex function of gas flow, source and wire temperature, as well as wire surface properties. However, as long as these are constant for the dataset which a fit is applied to, the complexities will be absorbed into the fit parameter.

Due to the parameter degeneracy within $A$, no determination of the total atom flow, $\Phi_\mathrm{at} = \Phi_\mathrm{mol} \cdot 2\alpha_{\rm{dissoc}}$, produced by the source will be attempted in this paper. Efforts presented here are focused on determining the relative angular output distribution of the source $j_{\rm{HABS}}(\theta ; \leff)$.

The recombination intensity can be expressed using $A$ as

\begin{equation}  
\label{eq:I_rec_short}
    I_\mathrm{rec} = A \cdot j_{\rm{HABS}}(\theta ; \leff)
    .
\end{equation}


Using the definition for beam intensity given in the equation above, a parametric model for the heating power due to recombination measured by the wire calorimeter can be defined. This takes the form of integrating $I_\mathrm{rec}$ over the surface of the wire:

\begin{align}
\label{eq:P_hit_fit}
{}&P_\mathrm{rec}(z_{\rm{pos}} ; \leff, A) = \nonumber\\
 {}&A \left( \int_{wire} dx \, j_{\rm{HABS}}(\theta(x, z_{\rm{pos}}); \leff)
 \cdot \eta(x) \cdot \cos^3(\theta(x, z_{\rm{pos}}) \right)
\nonumber\\ , \,
\end{align}
where $\eta$ is the relative sensitivity of the wire to heat depending on location (see Appendix \ref{sec:wire_sensitivity}), and $x$ is the position along the length of the wire over which the integration is performed. $P_\mathrm{rec}$ is evaluated as a function  of $z_{\rm{pos}}$, the position along the z-axis at which the wire is placed. $\theta$ can be calculated as a function of the Cartesian coordinates ($x,y,z$) per their definition shown in Appendix  \ref{sec:Axes Definition}. The multiplication with $\cos^3(\theta(x, z_{\rm{pos}}))$ results from the transformation between Cartesian and spherical coordinates.

This yields a model for the signal expected on the wire detector, which  can be fit to data as will be shown later. It has two principal parameters: the scaling parameter $A$ and $\leff$ which describes the shape of the distribution.

\section{Calibration}
\label{sec:Calibration}

\begin{figure*}[htbp]
    \centering
    \begin{subfigure}[t]{0.5\textwidth}
        \label{fig:Calibration_P_over_R}
        \centering
        \includegraphics[width=1\linewidth]{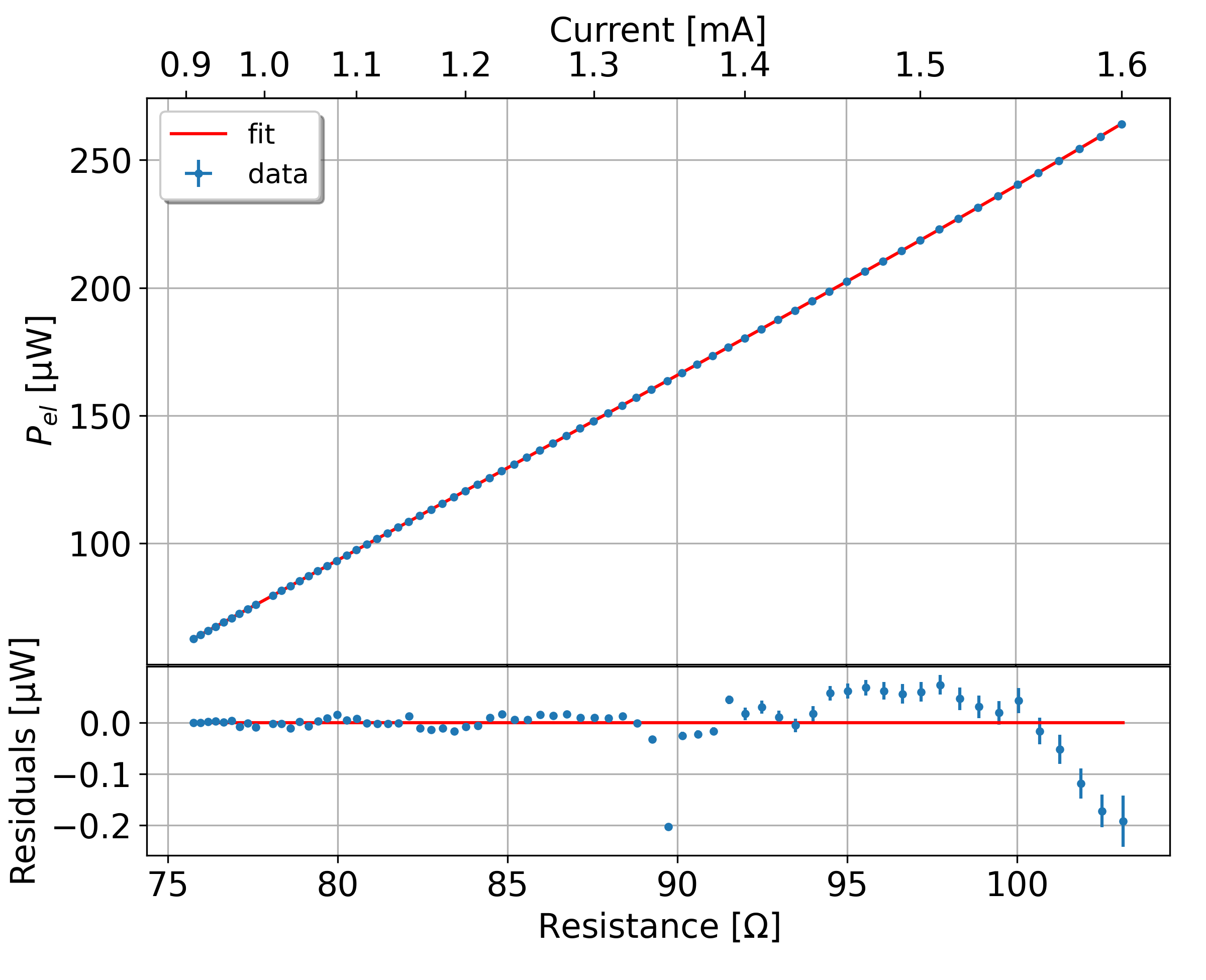}
        \caption{Electrical power supplied to the wire plotted over the resistance the wire reaches when heated with that power. The data is fitted with a 4th order polynomial. The data deviates from the fit function used towards the upper end of the range measured; however deviations are never more than 1.5 per thousand and so are accepted in the current analysis. The significant deviation in the data point at 90 $\Omega$  is unexplained, and we treat it as an aberration. It has no significant effect on the fit. 
        }
    \end{subfigure}%
    ~ 
    \begin{subfigure}[t]{0.5\textwidth}
        \label{fig:Calibration_k_overR}
        \centering
        \includegraphics[width=1\linewidth]{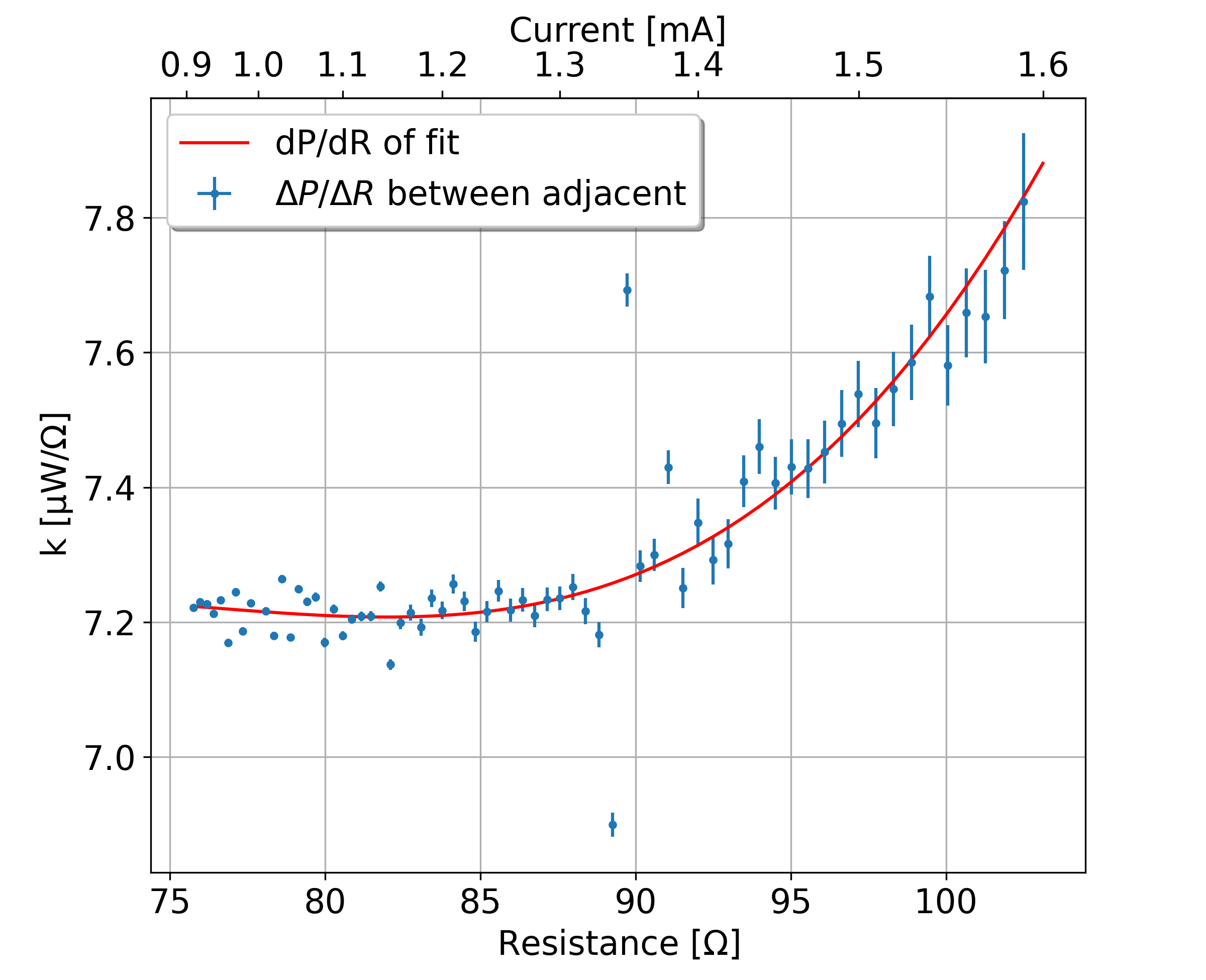}
        \caption{Calibration function k extracted from the slope of the power-resistance curve. The local approximation of the slope $\Delta P/\Delta R$ is calculated between adjacent data points and shown to illustrate the  scatter. The outlier at 90 $\Omega$ causes the local approximations to fluctuate significantly, but the fit curve which is used for further analysis suppresses this effect.
        }
    \end{subfigure}
    \caption{Calibration curve produced by varying electrical power supplied to the  wire and measuring the resulting wire resistance. The measurement is realized by changing the current across the wire from $0.9\,\rm{mA}$ to $1.6\,\rm{mA}$ in $0.01\,\rm{mA}$ steps. The resulting range of wire resistances corresponds to the range observed when heating the wire with external sources, primarily the thermal radiation from the HABS when heated to 2200 K.}
    \label{fig:Calibration}
\end{figure*}

In order to convert between changes in wire resistance and the heating power causing them, a
calibration of the detector is performed by running an increasing series of currents over the  wire. Since input voltage ($V$) and  current ($I$) through the wire are independently measured, both the resistance, as well as a direct measurement for the electrical power imparted to the wire via Joule heating ($P_\mathrm{el} = V \cdot I$) is performed simultaneously.
 
To produce a calibration function, a 4th order polynomial function of resistance is fit to the calibration data as shown in Figure \ref{fig:Calibration}. A 4th order polynomial is chosen because the highest-ordered effect is that of thermal radiation ($f_{\mathrm{rad}}$ defined in Section \ref{sec:Signal components}), which scales with  the 4th power in temperature, and  because it empirically fits the data to about 1 part in one thousand over the resistance range used in our measurements. This means potential systematic calibration errors due to the parametrization of this fit are significantly smaller than the statistical errors on beam measurements of about 1\% in the best case (see Figure \ref{fig:1sccm_signal_extraction}).

The calibration function $k$ can then be expressed as the derivative of the fit function
\begin{equation}
\label{eq:k}
k(R') = \left. \frac{dP}{dR}\right|_{R'} 
,
\end{equation}
such that
\begin{equation}
\label{eq:Power_uncorrected}
\Delta P_{\rm{calib}}(R',\Delta R) = \left. \frac{dP}{dR}\right|_{R'} \cdot \Delta R,
\end{equation}
where $\Delta P$ is the change in heating power supplied to the wire required to produce a small change in measured wire resistance $\Delta R$ away from an initial resistance $R'$. The choice of local calibration around an initial $R'$ rather than a global calibration is taken, as the initial resistance is subject to changes due to background effects, for example due to a change in the temperature of the surrounding apparatus. In practice the changes in resistance due to the atomic beam are always small, so the implicit locally linear approximation made when using the local slope approximation is justifiable.

When taking data with an external heat source like the beam, we additionally need to account for the feedback Joule heating due to the measurement current.

\begin{equation}
P_{fb} = I^2 \cdot \Delta R.
\end{equation}

With a constant measurement current $I$, an externally caused resistance change $\Delta R$ will result in additional electrical power draw and therefore heating of the wire, which needs to be subtracted to arrive at the measured external heating $P_{meas}$



\begin{align}
P_{meas}
    &=  \Delta P_{calib} - P_{fb}
    \\
    &=  k(R') \cdot \Delta R - P_{fb}
    \\
    \label{eq:Power_corrected}
    &=\left( k(R') - I^2\right)\cdot \Delta R
     .
\end{align}
With the typical operating current $I$ of 1 mA this results in a correction of $-1 \, \rm{\mu W / \Omega}$.

Finally an additional percent-level correction based on the fact that electrical heating power is non-uniformly distributed along the wire due to the temperature gradient is applied. Details about the derivation are discussed in Appendix  \ref{sec:correcting calibration}.

\FloatBarrier

\section{Signal components} 
\label{sec:Signal components}

In this section we will discuss a finite-element thermal model of the wire which illustrates the scaling of various heat fluxes on the temperature - and by extension resistance - of the wire, although it is not ultimately used as an input to our data analysis. The simulation is realized as Python code available on GitHub\footnote{\href{https://github.com/christianmatthe/wire_detector}{\url{https://github.com/christianmatthe/wire_detector}}}.

 The wire is divided into a number of segments of equal length and the heat flow $f$ between these elements is computed with the following equation (for each time step):
\begin{eqnarray}
\label{eq:f_tot}
    f_\mathrm{tot} &=& f_\mathrm{el} + f_\mathrm{rec} + f_\mathrm{beam\_gas} + f_\mathrm{bb}  \nonumber\\
    &&- f_\mathrm{ rad} - f_\mathrm{conduction}-f_\mathrm{bkgd\_gas},
\end{eqnarray}

This is the heat flow per unit length ($[f]=\rm \frac {Power}{Length}$),
where $f_\mathrm{el}$ is due to electrical current,
$f_\mathrm{rad}$ is due to radiation (emission from the wire and absorption from the surrounding chamber),
$f_\mathrm{conduction}$ is due to thermal conduction along and out of the wire,
$f_\mathrm{beam\_gas}$  is due to the kinetic energy of atoms and molecules in the beam,
$f_\mathrm{rec}$ is due to the recombination of hydrogen atoms on the surface of the wire, $f_\mathrm{bb}$ is due to the blackbody radiation from the glowing hot HABS, 
and $f_\mathrm{bkgd\_gas}$ is due to interactions with the background gas in the chamber.
The equations describing each individual component can be found in Appendix  \ref{sec:Wire Heating component equations}.

Integrating any of these heat flows $f_\mathrm{x}$ over the length of the wire yields $P_\mathrm{x}$, the total heating (cooling) power due to this source (sink).

 $f_\mathrm{rec}$ contains information about the amount of atomic hydrogen that is in the measured beam and is therefore the component of interest for this paper. The procedure for extracting it from  the raw signal is detailed in the following subsections.

From the net heat flow the temperature change per time step is calculated 
\begin{equation}
\Delta T = \frac {f_\mathrm{tot} \cdot  \Delta t}{\rho \cdot A \cdot c}
\end{equation}
where $\rho$ is the density of the wire material, $A$ is the cross sectional area, $c$ is the specific heat capacity of the material, and $\Delta t$ is the length of one time step in the simulation. The simulation is run for a number of steps, until $\Delta T$ converges to 0, i.e. thermal equilibrium. The  $\frac{1}{e}$ equilibration time constants $\tau$ are typically between $0.3\, \rm{s}$ to $0.5\,\rm{s}$ depending on the precise parameters used, so we run simulations for at least ten simulated seconds to ensure full equilibration. The resistance of the wire can then be calculated from the resulting temperature distribution, given the temperature-dependent resistivity coefficient of the wire material.

\begin{figure}[htbp]
    \centering
    \includegraphics[width=1\linewidth]{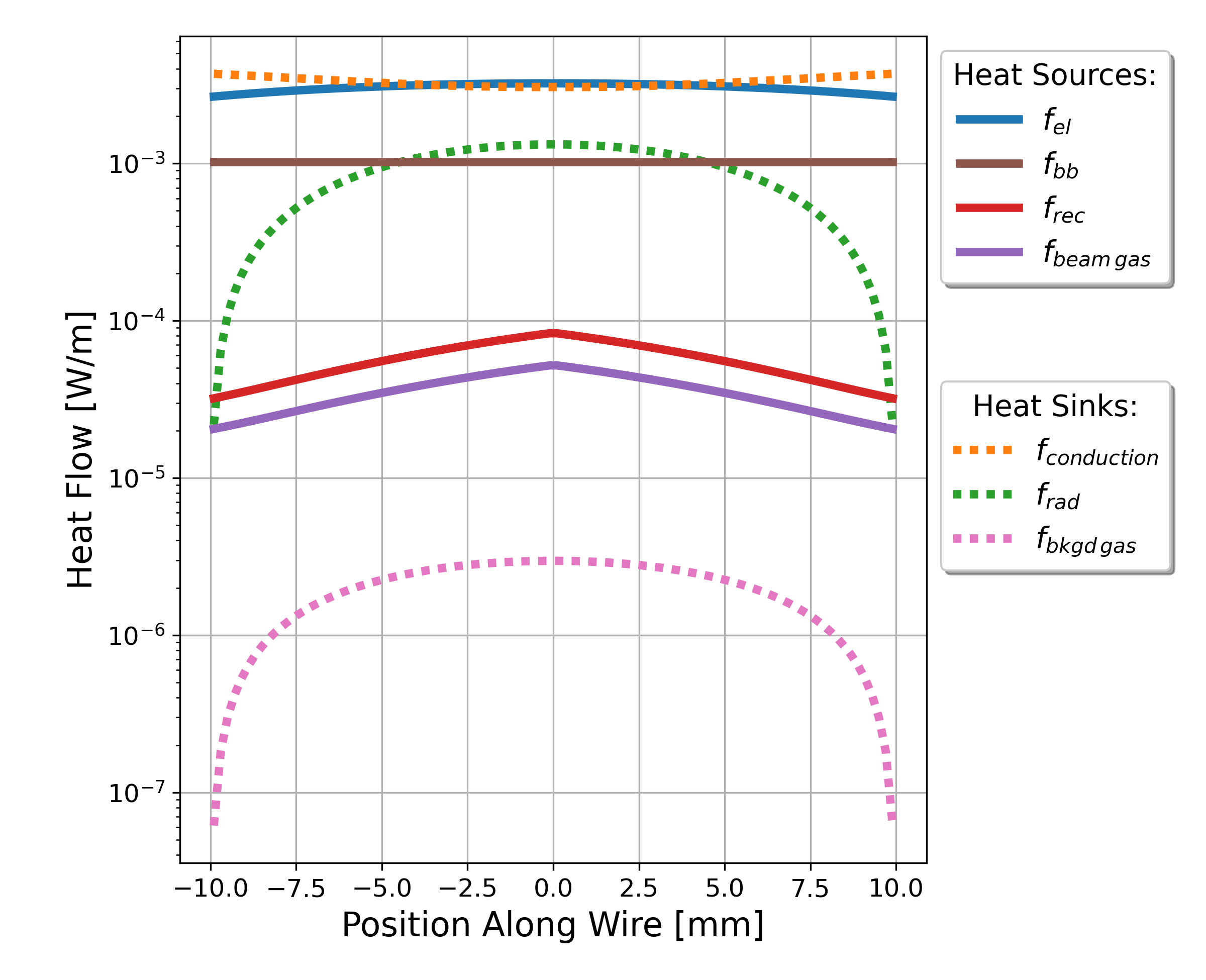}
    \caption{
    Comparison of the scale of the various simulated heat flow components at the time of equilibrium over the extent of the wire. Sources which transfer heat into the wire are plotted with solid lines, while sinks that remove heat from the wire are drawn in dashed lines. Note that the heat sinks are subtracted in Eq. \eqref{eq:f_tot}, and therefore cancel out a heat source of equal size in this plot. The hydrogen beam selected for this simulation contains 1 sccm of molecular hydrogen at 2200 K with a 10\% dissociation rate into hydrogen atoms. The beam width is chosen to be consistent with beams at this flow rate as will be measured later in this paper.
    }
    \label{fig:Heat_flow}
\end{figure}

Figure \ref{fig:Heat_flow} shows the heat flows after equilibrium has been reached for typical settings for gas flow (1 sccm) and HABS temperature (2200 K). It illustrates the relative size of the effects impacting the wire temperature. Most important to note is, that both the $f_\mathrm{el}$ and $f_\mathrm{bb}$ are significantly larger than the heat sources $f_\mathrm{rec}$ and $f_\mathrm{beam\_gas}$ caused by a typical beam. This means that both must be extremely well-stabilized to be able to see the beam signal among small fluctuations in larger effects.

 Fluctuations in $f_\mathrm{el}$ are minimized by using a highly stable power supply~\footnote{Keithley SourceMeter 2400}, capable of stabilizing the 1 mA measurement current running across the wire to better than $10^{-5}$ relative fluctuations. To minimize fluctuations in $f_\mathrm{bb}$ the HABS is operated on a PID-controlled loop stabilizing the readout of an internal thermocouple in close proximity to the HABS capillary.

However, it has not been feasible to model all effects impacting the wire with sufficient precision to formulate a relationship for extracting beam heating power from wire resistance directly. In part this is due to the great number of material and surface properties that enter into the model, many of which are only available with insufficient precision. Even the ostensibly simple calculation of the wire's baseline resistance at room temperature turns out to be problematic, because the manufacturer only guarantees the diameter of the wire to a 10\% precision. This is further complicated because a gold-coated tungsten wire is used, where the exact impact of the coating on wire resistance is difficult to model with great confidence. Exact measurements of the sticking fractions and accommodation coefficients for $f_\mathrm{beam\_gas}$  and 
$f_\mathrm{rec}$ are very challenging to determine with sufficient precision \cite{Melin1970}.
Additionally, changes in chamber temperature not modeled here also cause changes in wire resistance larger than that of the expected signal.

Ultimately, this simulation is used to inform the interpretation of the data but not to directly derive a beam signal from wire resistance. Instead an indirect extraction of the signal  (Section \ref{sec:signal extraction}) is used along with a calibration scheme (Section \ref{sec:Calibration}) for correlation of wire resistance to incident power measured.

\section{Signal extraction}
\label{sec:signal extraction}
\subsection{Primary signal extraction} 
\label{sec:primary signal extraction}

\begin{figure}[htbp]
\centering
	\includegraphics[width=1\linewidth]{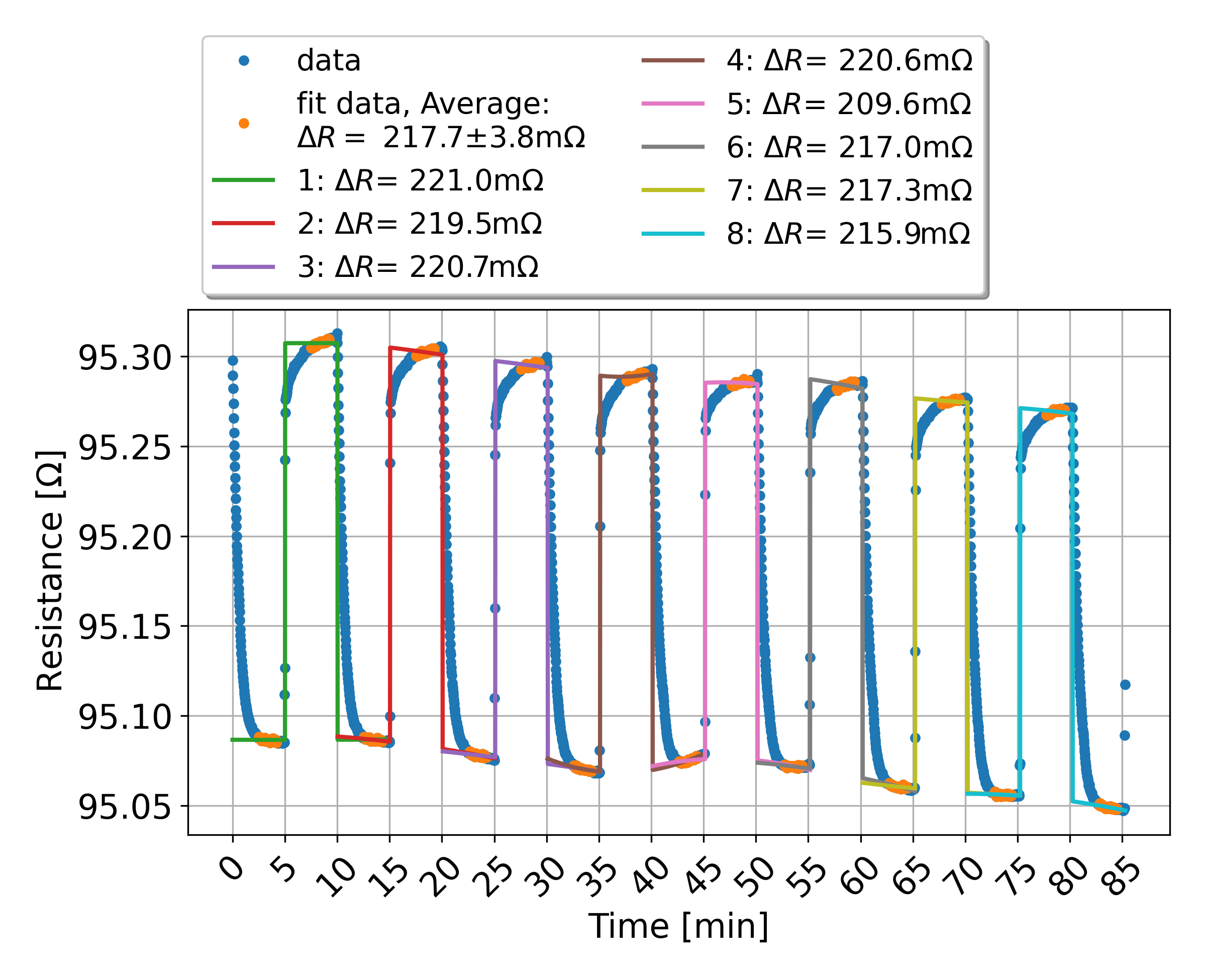}
    	\caption{
     Example of signal extraction for the wire positioned centrally in the beam at $z = 1.5 \rm{mm}$, with the HABS at $\approx 2200 \, \rm{K}$ and a flow of 1 sccm. The flow is cycled on and off 8 times. For each cycle the difference in resistance when the flow is on is extracted using a fit to a subset of the data after equilibration, marked in orange. Fits are applied individually to every set of three 5 minute segments in an beam off-on-off pattern.
     }
	\label{fig:1sccm_signal_extraction}
\end{figure}

The beam heating effects on the wire ($f_\mathrm{beam\_gas}$ and
$f_\mathrm{rec}$) can be quite small, on the order of $10^{-9} - 10^{-6} \,\rm{W}$. Since the wire is subject to external temperature drifts, signals are extracted as a relative measurement by periodic switching between the "beam on" and "beam off" states, to allow for constant comparison against the drifting background. An example of this is shown in Figure \ref{fig:1sccm_signal_extraction}, where a typical hydrogen beam is cycled on and off every 5 minutes by opening and closing a pneumatic valve in the gas supply. For each 5 minute section only the last two minutes of data are used to allow for equilibration. This data is marked orange in the plot. Then three 5 minute sections are combined in an "off"-"on"-"off" pattern. The two "off" sections allow for fitting the background trends and then determining the average difference $\Delta R$ when the beam is "on" in the central 5 minute section. For background subtraction, a quadratic function is fit to the data, with an additional $\Delta R$ offset factor only applied to the "beam on" data.

\begin{equation}
R_{\rm{fit}}(t) =   
\begin{cases} 
c_0 + c_1 t + c_2 t^2 & \textrm{if \quad "beam off"} \\
c_0 + c_1 t + c_2 t^2 + \Delta R & \textrm{if \quad "beam on"} \\

\end{cases}
,
\end{equation}

Tracking the change in resistance in the "beam off" state, as the flow switching method does, practically eliminates any slow-changing background effects. The measurement turns into a comparison of two  values, where the only variable is hydrogen flow through the HABS. This eliminates the need to model effects such as $f_\mathrm{el},f_\mathrm{bb} $ which are not affected by hydrogen flow directly with great precision in order to extract the much smaller beam-related effects.

We do not attempt to model the equilibration process of each cycle, as it is a potentially complex superposition of thermalization processes of various components (wire, detector board, vacuum chamber), as well as the equilibration of gas flow after switching, all of which take place over different time scales. From the simulation and measurements not presented here, it is known that the wire temperature equilibrates on the scale of a few seconds, while the detector PCB temperature has been measured to take tens of minutes to reach full equilibrium. The 5 minute cycle is a compromise in letting the fast processes that affect the wire directly equilibrate, while not waiting for all components to fully thermalize over e.g. one hour, because the ability to background-subtract unintended external temperature drifts would be lost. The partial inclusion of background temperature equilibration is undesired but not easily avoidable. It represents a systematic overestimation of the true signal on the order of 20\%. Since our model in Eq. \eqref{eq:P_hit_fit} includes a scaling parameter $A$, for the purposes of this paper this systematic will simply be absorbed by $A$. This is yet another reason why an absolute calibration is difficult to achieve, but the effects on measuring the beam shape in terms of relative flux, as shown  in Section \ref{sec:Application to Data}, should be minor.

The flow switching method enables checking for repeatability and stacking of multiple measurements to increase SNR. 
 With 8 such repetitions, signals as small as $ \Delta R = 2 \,\mathrm{m \Omega}$ (equivalent to $\approx 13 \,\mathrm{nW}$, or $\approx 10^{14} \, \mathrm{atoms/(cm^2 s)}$) have been resolved with an SNR of $\dfrac{\mu_{\Delta R}}{\sigma_{\Delta R}}\approx 4$. 

After $\Delta R$ is extracted, we use Eq. \eqref{eq:Power_corrected} to calculate $P_\mathrm{meas}$, the net heating power responsible for the resistance change.

\subsection{Secondary signal extraction}
\label{sec:secondary_extraction}
The result of primary signal extraction is the net heating power as derived from the net change in resistance. In a second extraction step, the atomic hydrogen recombination heating $f_\mathrm{rec}$ must be extracted from the net heating power.

The assumption is made that the components remaining in the signal are $P_\mathrm{meas} = P_\mathrm{rec} + P_\mathrm{beam\_gas} - P_\mathrm{bkgd\_gas}.$ The other components listed in Eq. \eqref{eq:f_tot} ought to be eliminated by the flow switching scheme of the primary signal extraction.

With the HABS actively putting gas into the vacuum chamber the background pressure is typically $\simeq 2 \times 10^{-5} \, \rm{mbar}$ at the 1 sccm flow that is used for measurements presented in this paper. Assuming that the background gas component is in thermal equilibrium with the walls and has no atoms in it due to the recombination and thermalization on the chamber walls, it will provide a constant cooling effect. The hot molecules in the beam in contrast will deposit power increasing with temperature. One source of non-linearity that needs to be accounted for is the non-trivial scaling of the heat capacity of molecular hydrogen with temperature\footnote{NIST \url{https://webbook.nist.gov/cgi/cbook.cgi?ID=C1333740&Type=JANAFG&Table=on#JANAFG}} $C_{V, \rm{H_2}}(T)$.

To extract just the power due to recombination, equivalent measurements are taken at three HABS temperature setpoints $T_{low}, T_{mid}, T_{high}$. $T_{low}$ is chosen as room temperature, $T_{mid}$ at $\approx 1277\,\rm{K}$, high enough to see  the effects of $f_\mathrm{beam\_gas}$ but low enough that no atomic hydrogen will be produced, and $T_{high}$ is chosen to be $\approx 2211\, \rm{K}$, the highest reliably reachable temperature with the hardware employed. The precise numerical values of the temperatures seem arbitrary, because they are the conversion of flat thermocouple voltages which are used for stabilization. While the stabilization is quite good,  the absolute conversion should be taken as no more than 5\% accurate, as precise calibration of capillary temperature measurements remains a challenge.

The measurements at $T_{low}$ and $T_{mid}$ are the minimum number of points required to scale the effects of  $f_\mathrm{beam\_gas} - f_\mathrm{bkgd\_gas}$ to what they should be at $T_{high}$. The scaled effect is subtracted from the raw measurement to obtain the extracted recombination heating power

\begin{align}
\label{eq:P_rec}
 P_{rec}(T_{high})  = & P_{meas}(T_{high}) - (a \cdot C_{V, \rm{H_2}}(T)\cdot T_{high} + b) \, , 
\end{align}
where $P_{meas}$ is the heating power extracted via Eq. \eqref{eq:Power_corrected}, as described in Section  \ref{sec:Calibration}, at each of the temperature setpoints. $a$ and $b$ are determined by fitting  
\begin{equation}
\label{eq:P_H2_fit}
P_{\rm{H_2}} = (a \cdot C_{V, \rm{H_2}}(T)\cdot T + b)
\end{equation}
 to the $T_{low}$ and  $T_{mid}$ measurement for every $z_{\rm{pos}}$ the  wire is placed at.
This amounts to an extrapolation and subtraction of the non-recombination background effects, primarily heating from hot but undissociated $\rm{H_2}$ molecules in the beam.

\begin{figure}
    \includegraphics[width=1\linewidth]{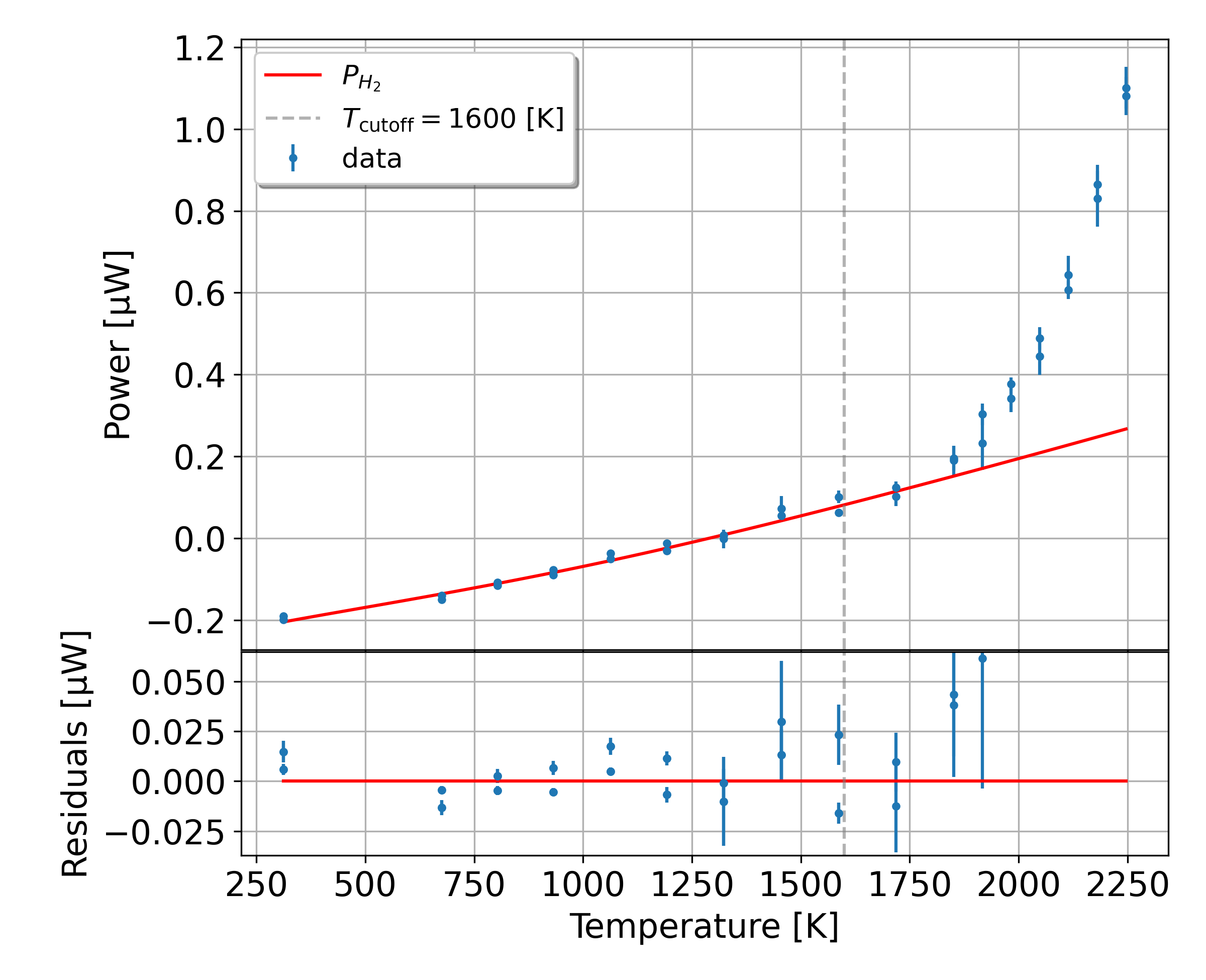}
    \caption{Power detected on the wire after primary extraction ($P_{\rm{meas}}$) at $1 \, \rm{sccm}$ for a fixed z-position over a range of temperatures. The temperature scan is performed twice in this dataset for each temperature position, once scanning up in temperature and once down. A fit based on Eq. \eqref{eq:P_H2_fit} is applied to data points below the temperature cutoff $T_{\rm cutoff} = 1600 \, \rm K$. Below this temperature no dissociation of hydrogen is expected, and therefore no contribution of atomic hydrogen  to the heat detected on  the wire.}
    \label{fig:plot_fit_H2}
\end{figure}

Figure \ref{fig:plot_fit_H2} illustrates such a fit using a dataset with higher resolution in HABS temperature at a single wire z-position. The temperature scan is performed twice at each temperature, once scanning up in temperature once going  down. In the example, Eq. \eqref{eq:P_H2_fit} is applied to the data below a temperature cutoff of 1600 K to ensure none of the measured power is due to atomic H. The model describes the data below the cutoff well, but with discrepancies not fully explainable by statistic fluctuation. These may be partially the a result of imperfect calibration of the temperature axis. Work on improved temperature determination is ongoing, but not presented in this paper.
Beyond that, Eq. \eqref{eq:P_H2_fit} may not fully capture all temperature dependencies if further parameters in  Eq. \eqref{eq:f_beam_gas} carry significant temperature dependence. For example, previous work we have carried out, but not yet published, suggests a limit of the temperature dependence of $\alpha_{E_{mol.}}$ of at most 5\% in our temperature range.

A choice of only three representative temperature points is made to minimize total required data taking time. 
Since full z-scans of the resolution presented in Figure \ref{fig:multi_run_power_subset_1sccm} require 80 hours of data taking per temperature, using high resolution in temperature as well quickly becomes operationally prohibitive. Based on the dataset presented in Figure \ref{fig:plot_fit_H2} we estimate the three temperature point method introduces roughly an additional 5\% uncertainty on the extracted $P_{\rm rec}$ compared to doing a full resolution fit when the baseline between the chosen $T_{low}$ and  $T_{mid}$ is at least 500 K.

Eq. \eqref{eq:P_rec} describes the procedure by which measurements at multiple temperatures can be combined to extract only the heating due to hydrogen recombination. This recombination heating is used in this work to trace the relative flux density of atomic hydrogen. Therefore the preceding section has described the prescription for using a wire calorimeter as a detector for atomic hydrogen, so long as the assumptions that were made are valid.

\section{Application to data at 1sccm}
\label{sec:Application to Data}

\begin{figure}[H]
    \centering
    \includegraphics[width=1\linewidth]{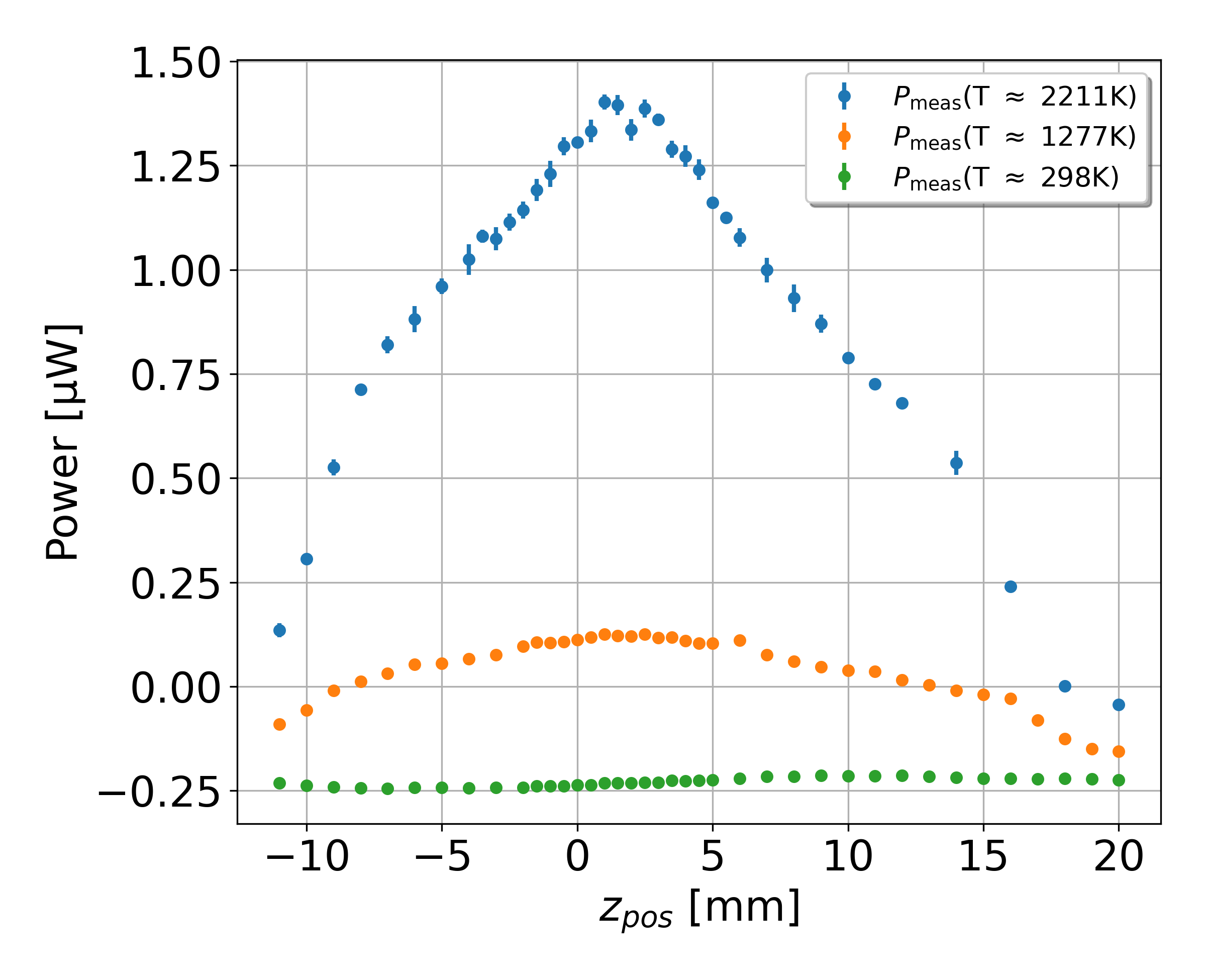}
    \caption{Power detected on the wire after primary extraction ($P_{\rm{meas}}$) at $1 \, \rm{sccm}$ for room temperature, a non-dissociating medium temperature, and a high temperature where atom production is expected. These data sets are then used in the secondary extraction procedure, to yield the effect of recombining atoms on the surface. Negative detected powers indicate a net cooling of the wire, which is identified with the cooling due to gas pressure increase in the detector chamber induced by gas flow from the source. ($f_\mathrm{bkgd\_gas}$). }
    \label{fig:multi_run_power_subset_1sccm}
\end{figure}

As a demonstration, the methods described in the previous sections are applied to data taken by sweeping across the hydrogen beam with the wire detector by means of the translation stage on which it is mounted. Figure \ref{fig:multi_run_power_subset_1sccm} shows data taken when flowing 1 sccm  ($4.48 \times 10^{17} \,\rm{ molecules/s}$) of hydrogen through the HABS at three capillary temperatures $T_{low}\sim298\,K$, $T_{mid}\sim1277\,K$ and $T_{high}\sim2211\,K$. The range of z-positions represents the full range available in the limited space of the vacuum chamber.
To avoid an impact of time-dependent effects over the measurement period of many hours, data at the different positions are acquired in non-consecutive order.

As seen in Figure \ref{fig:multi_run_power_subset_1sccm}, room temperature measurements initially yield a negative net heating power. This is because at this low temperature the beam imparts no significant heating to the wire, but it still fills the vacuum chamber with gas such that the pressure rises. This room temperature background gas cools the wire, since the wire reaches $\approx 70\degree\,\rm{C}$ in the center due to being heated by the measurement current. Both effects are described in Section  \ref{sec:Signal components}. 

At medium temperatures ($1277\,\rm{K}$ in Figure \ref{fig:multi_run_power_subset_1sccm}) there is a slight heating on the order of $300\,\rm{nW}$ which can be seen to be higher when the wire is centered under the source ($z_{\rm{pos}} \approx 1.5\,\rm{mm}$). This can be accounted for by a beam of hot molecular $\rm{H}_2$ which transfers some of its thermal energy to the wire when hitting it ($f_\mathrm{beam\_gas}$).

Finally at high temperatures ($2211\,\rm{K}$ in Figure \ref{fig:multi_run_power_subset_1sccm}) larger than the dissociation threshold temperature, the beam contains a fraction of atomic hydrogen. This is the primary signal we are looking for, and can account for the much larger µW-scale peaked signal in the high temperature power data plotted in blue. The high temperature dataset shows a larger signal than can be explained by the increased thermal "beam gas" heating indicating the detection of recombination heating due to atomic hydrogen. 

\begin{figure}[htbp]
    \centering
    \includegraphics[width=1\linewidth]{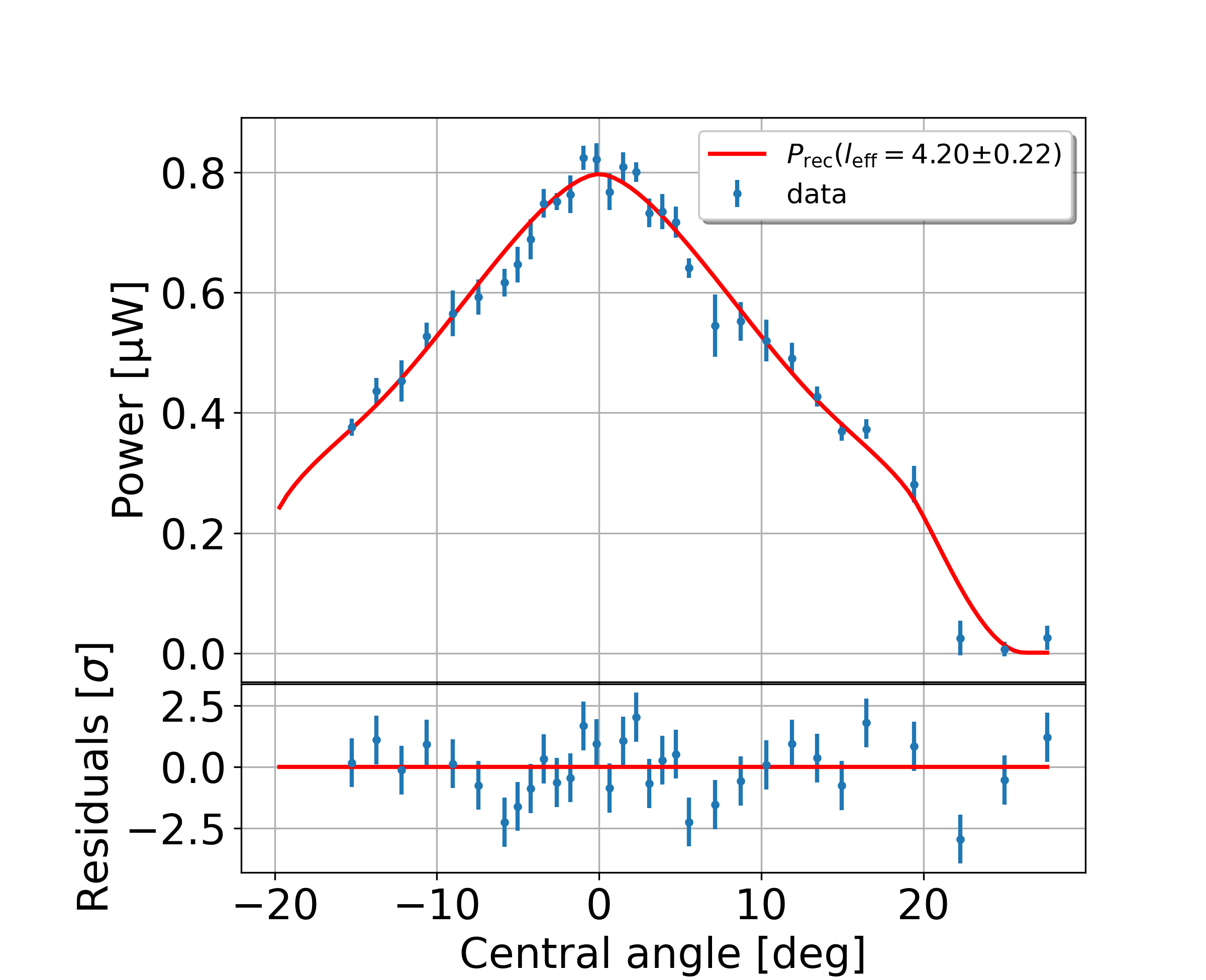}
    \caption{Extracted heating power attributed to recombination ($P_{\rm{rec}}$) for a flow of $1 \, \rm{sccm}$ at $2211 \,\rm{K}$ source temperature, shown over the range of positions the wire was placed at. The optimized beam shape model is plotted alongside. 
}
    \label{fig:fit_1sccm_2250K_penumbra}
\end{figure}

These three data sets are combined using the procedure described in Section  \ref{sec:secondary_extraction} using Eq. \eqref{eq:P_rec} to determine the heating power due to recombination heating alone. This is plotted in Figure \ref{fig:fit_1sccm_2250K_penumbra}.

The model from Eq. \eqref{eq:P_hit_fit} is then adapted for fitting to the extracted recombination power data. An offset power $P_0$, to account for a residual background, as well as a positional offset $z_0$ are introduced. The necessity of this can be seen in Figure \ref{fig:multi_run_power_subset_1sccm}, where the zero of  the z-axis does not initially line up with the center of the beam. The resulting final fit function is given by

\begin{align}
\label{eq:P_hit_fit_long}
{}&P_\mathrm{rec}(z_{\rm{pos}} + z_0 ; \leff, A, P_0) = \nonumber\\
{}&A \left( \int_{wire} dx \, j_{\rm{HABS}}(\theta(x, z_{\rm{pos}} + z_0))
 \cdot \eta(x) \cos^3(\theta(x, z_{\rm{pos}} + z_0) \right) \nonumber\\
{}&+ P_0 \quad . \quad
\end{align}

Figure \ref{fig:fit_1sccm_2250K_penumbra} shows the fit applied to the final extracted data. As previously described, the parameter $A$ is effectively just a vertical scaling parameter, while $P_0$ and $z_0$ translate the fit in x and y direction. 
This leaves $\leff$ as the sole free parameter governing the shape of the beam's distribution.

\begin{figure}[htbp]
    \centering
    \includegraphics[width=1\linewidth]{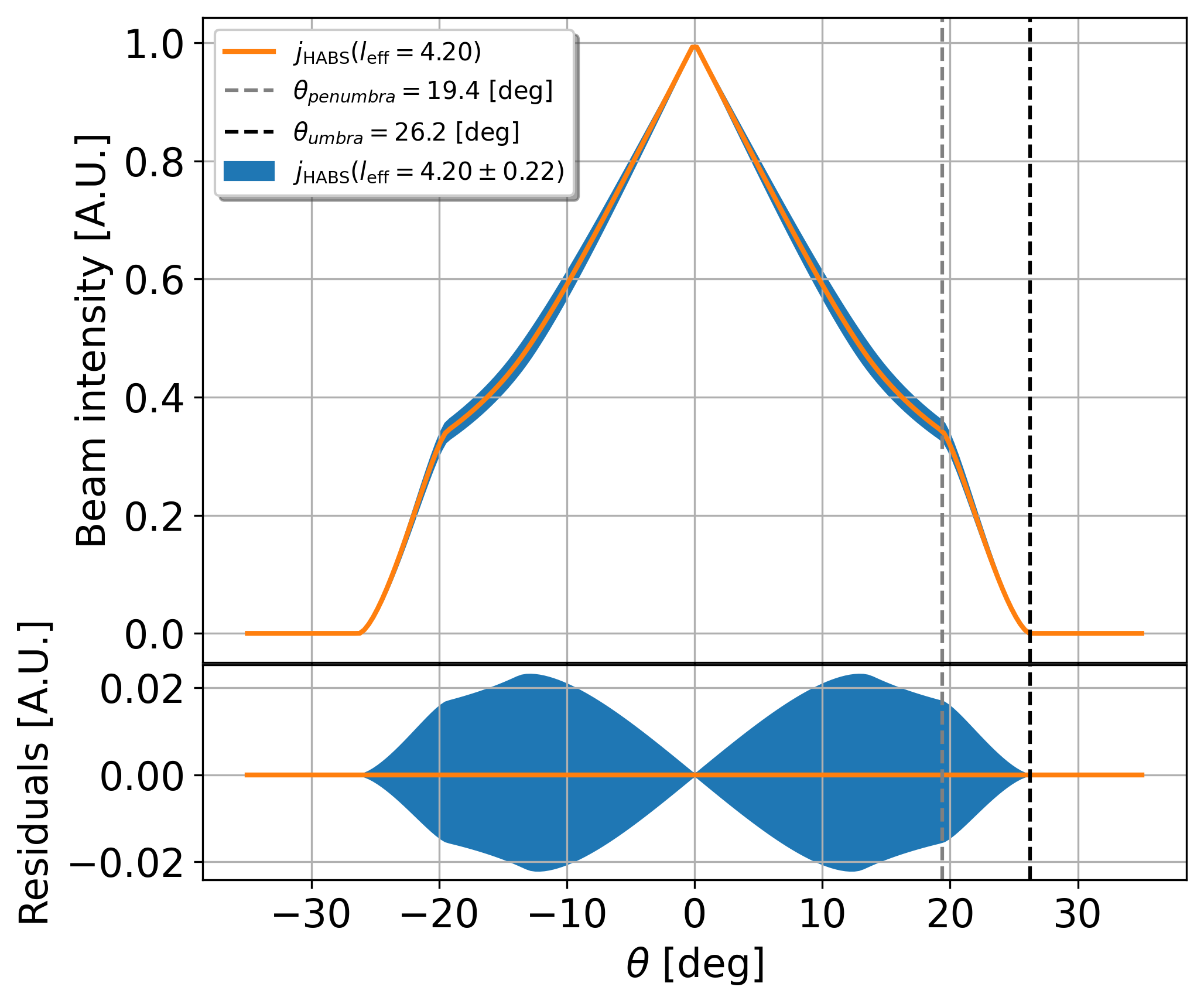}
    \caption{Reconstruction of the underlying beam shape model $j_{\rm{HABS}}$, including both the effects of the theoretical beam shape emitted from a long tube, as well as the truncation of the edges due to shadowing of the beam by the source shroud, both as described in Section  \ref{sec:Beam source}. The uncertainty on the extracted beam model is illustrated by propagating the fit errors for the beam shape parameter $\leff$ through the model. The errors on $\leff$ allow for a reconstruction of the beam as it exits the HABS to better than 5\% relative precision at any point along the distribution.}
    \label{fig:H_profile_penumbra}
\end{figure}

This fit yields a value for the shape parameter $\leff$, which allows us to reconstruct the beam intensity profile at the exit of the HABS.
The resulting $j_{\rm{HABS}}$ as per Eq. \eqref{eq:H_profile} is shown in Figure
\ref{fig:H_profile_penumbra} alongside an error band. The error band results from calculating $j_{\rm{HABS}}(\leff)$ for the values $\leff$ shifted up and down by the fit error on $\leff$. The error band deviation is zero at the center and at the edges by construction of the model, since it only shows relative flux. Any changes in flux scaling (errors in $A$) and background offset (errors in $P_0$) are external to this model. They would be important for a measurement of absolute flux, but not relative flux as is presented in this work.
This example demonstrates that the procedure presented allows for measuring the shape of the original beam to better than 5\% relative precision at any point along the distribution.

\section{Effective length over flow}
\label{sec:leff_over_flow}
\begin{figure}
    \centering
    \includegraphics[width=1\linewidth]{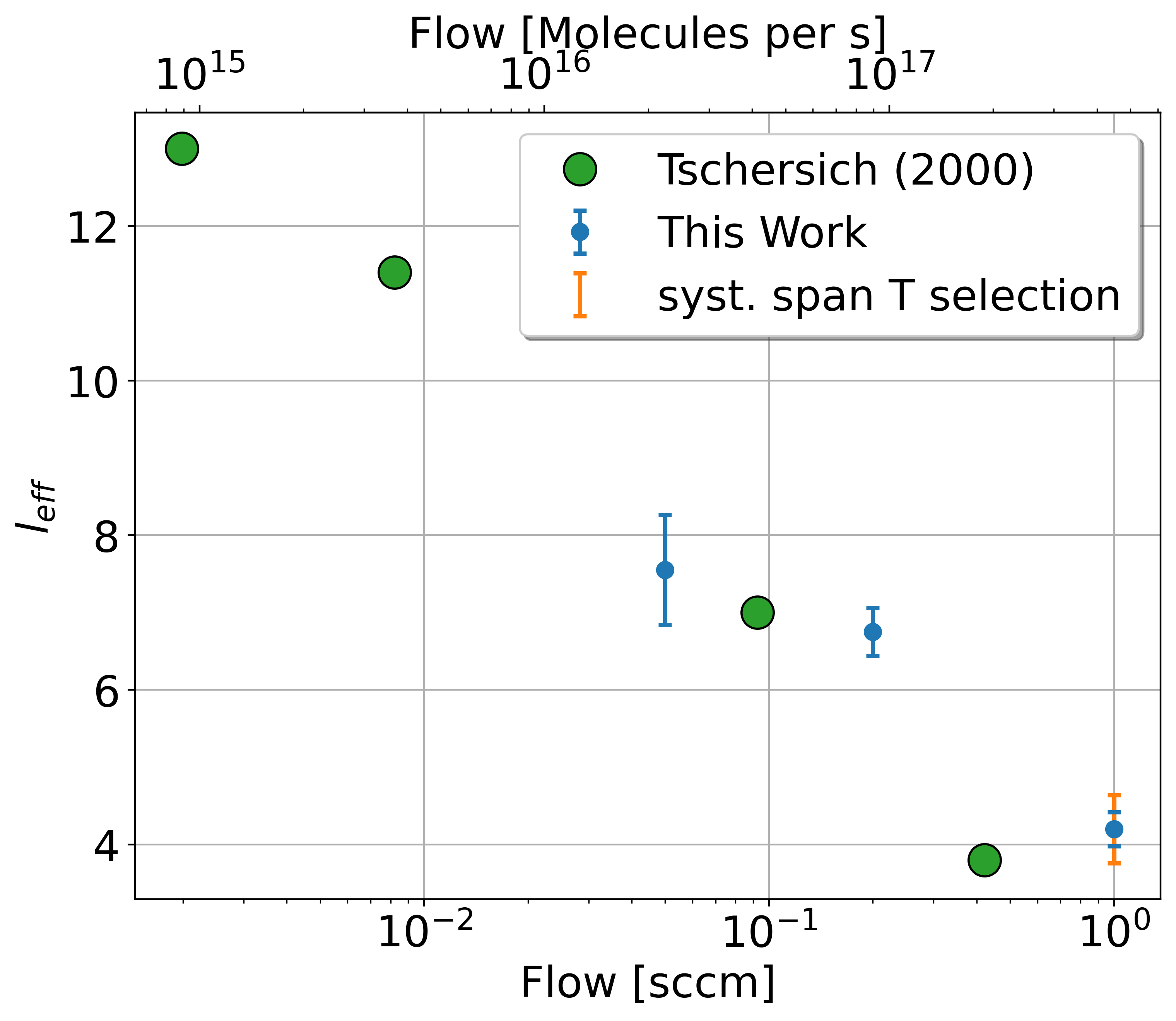}
    \caption{Comparison of fit parameter $\leff$ at three different flows using the methods described in this work to values from Table I in Tschersich et al. (2000) \cite{Tschersich2000}. No error estimation is available for the sourced $\leff$ values. Plots showing fits in this work for 0.05 sccm and 0.2 sccm are available in Appendix \ref{sec:Appendix_leff_over_flow}. The orange error bar is the span of $\leff$ when datasets at different temperatures are used to determine $P_{H_2}$ (see Appendix \ref{sec:systematics_T}) for details).
     }
    \label{fig:leff_over_flow}
\end{figure}

The same three temperature scan procedure as illustrated in Section \ref{sec:Application to Data} has been performed for beams at 0.05 sccm and 0.2 sccm flow. This provides overlap with the lower flow regimes probed by Tschersich et al \cite{Tschersich2000}. Values for $\leff$ listed in Table I of that work are compared to the ones produced by this work in Figure \ref{fig:leff_over_flow}. Extending the range of flows measured with the wire detector is the subject of ongoing work.

The two works provide values that are roughly compatible. In either case the effective length of transparent flow is measured to increase with lower flows, which matches expectations, as lower flows correspond to lower densities in the capillary. The exact quality of the comparison is hard to determine. Tschersich  et al. do not quote errors on their $\leff$. Additionally while their source is very similar to the source used in this work (Section \ref{sec:Beam source}) and indeed is a prototype of the same, it is not exactly equivalent, especially in the precise geometry of its housing. They also operate their source at higher temperatures, although we do not expect this to have a significant direct effect on $\leff$.


\section{Conclusion}

The performance of the wire detector was demonstrated by measuring the relative intensity profile of the beam from this source. An independent absolute intensity measurement is not yet attempted, due to an undetermined proportionality factor for the fraction of available energy from recombination that is transferred into the wire.
It has been demonstrated that a calorimetric wire detector can be used to measure hydrogen beams on the order of $10^{16} \,\rm{atoms}/{(\rm{cm^2}\rm{s})}$ with significant SNR ($\dfrac{\mu_{\Delta R}}{\sigma_{\Delta R}}\approx 50$).
It is possible to reconstruct the underlying shape of the hydrogen beam emitted from a capillary based on a theoretical model, showing that remaining uncertainties are reasonably constrained. Good agreement with the theoretical model for the relative distribution of the beam was demonstrated. 
This constitutes an independent confirmation of this model as well as the suitability of the calorimetric wire detector for further development in characterizing hydrogen beam sources. The beam shape parameters determined with the method described in this work are compared to those from measured for a similar source using a QMS by another group \cite{Tschersich2000} and are found to be compatible.

\FloatBarrier
\appendix
\appendix

\section{Wire heating component equations}
\label{sec:Wire Heating component equations}
This section is a glossary of equations used in the simulation of the wire detector. All quantities $f_x$ are power densities (power per unit length $[\rm{W/m}]$) due to various effects that impact the wire.

- $f_\mathrm{el}$: Heat due to electrical current.

    \begin{equation}
    \label{eq:f_el}
        f_\mathrm{el} = I^2 \cdot R(T_i) / l
    \end{equation}
where $I$ is the current, $R(T_i) = \rho\cdot l/A \cdot (1 + a)\cdot(T_i-T_{\rm{ref}})$ is the resistance of the segment $i$, and $l$ is the length of the segment. $\rho$ is the specific resistance, $A$ denotes the cross-sectional area of the wire and $a$ is the temperature coefficient of resistivity. $T_i$ is the temperature of segment $i$ and $T_{\rm{ref}}$ is the reference temperature at which $\rho$ is quoted.
    
- $f_\mathrm{rad}$: Heat transfer through radiation (emission and absorption from surroundings).
    
    \begin{equation}
    \label{eq:f_rad}
        f_\mathrm{rad} = \pi\cdot  d_\mathrm{wire}\cdot  \sigma\cdot \epsilon \cdot (T_i^4 -T_\mathrm{bkgd}^4)
    \end{equation}
where $d_{\rm{wire}}$ is the diameter of the wire, $\sigma$ is the Stephan-Boltzmann-constant and $\epsilon$ is the emissivity of the wire surface. The emissivity is not precisely known from literature, and the value therefore has to be tuned using a calibration scheme. $\pi\cdot d\cdot l$ is the surface area, and divided by $l$ just results in $\pi \cdot d$. $T_\mathrm{bkgd}$ is the temperature of chamber which gives of a background of blackbody radiation.
    
- $f_\mathrm{conduction}$: Heat transfer through conduction.

    \begin{equation}
    \label{eq:f_conduction}
        f_\mathrm{conduction} =k(T_i)\cdot  \frac {( T_{i-1}-T_i)+(T_{i+1}-T_i)}{l}\cdot \frac Al
    \end{equation}
where $k$ is the heat conductivity coefficient and $A$ is the cross sectional area of the wire. This is the second-order central difference approximation, which calculates the local curvature in the temperature profile by comparing to the temperatures of the neighboring segments $T_{i-1}$ and $T_{i+1}$.
    
- $f_\mathrm{bkgd\_gas}$: Heat transfer due to interactions with the gas remaining in the vacuum chamber.
    

    \begin{align}
    \label{eq:f_background_gas}
        f_\mathrm{bkgd\_gas} &= \alpha_E \cdot \pi d_{\rm{wire}} \cdot \frac{p}{4} \sqrt{
        \frac {1}{k_B m T_\mathrm{bkgd}}} \\
        & \quad \cdot
        \left(\frac{C_V(T_i)T_i - C_V(T_\mathrm{bkgd})T_\mathrm{bkgd}}{N_A}\right) \notag \\
        &\approx  \alpha_E \cdot \pi d_{\rm{wire}} \cdot \frac{5p}{4} \sqrt{\frac {k_B}m\frac {(T_i - T_\mathrm{bkgd})^2}{T_\mathrm{bkgd}}}
        ,
    \end{align}
where $\alpha_{E}$ is the energy accommodation coefficient of molecular hydrogen on the wire, $p$ is the pressure in the chamber containing the wire detector, $k_B$ is the Boltzmann constant, m is the mass of a hydrogen  molecule, $C_V(T)$ is the heat capacity of the gas (molecular hydrogen in this case), and $N_A$ is the Avogadro number. $\alpha_{E}$ describes the fraction of available kinetic energy that is transferred to the collision surface and is set to 1 in the example shown in Figure \ref{fig:Heat_flow}.
    
- $f_\mathrm{beam\_gas}$:  The heat transfer from the kinetic energy of atoms and molecules in the beam hitting the wire.

    \begin{equation}
    \label{eq:f_beam_gas}
        f_\mathrm{beam\_gas} = f_\mathrm{at} + f_\mathrm{mol},
    \end{equation}

    \begin{equation}
    \label{eq:f_at_mol}
        f_\mathrm{at/mol} =  n_\mathrm{at/mol} \cdot d_{wire} \cdot \alpha_{E_\mathrm{at/mol}}\cdot C_{V_{\mathrm{at/mol}}}(T_\mathrm{at/mol})\cdot  (T_\mathrm{at/mol} - T_i)
    \end{equation}
where n is flow density $[\rm{particles}/\rm{(cm^2 s)}]$ and $\alpha_{E}$ is the energy accommodation coefficient and $T_\mathrm{at/mol}$ is the temperature of the gas coming our of the source. In the example presented in this paper the gas temperature is assumed to be equal to the source temperature. 

- $f_\mathrm{rec}$: Heat transfer due to the recombination of hydrogen atoms from the beam on the wire.
    
    \begin{equation}
    \label{eq:f_rec}
        f_\mathrm{rec} = j_{\rm{HABS}}(\theta(i);\leff)\cdot\Phi_{\rm{at}} \cdot \gamma_\mathrm{rec}\cdot \beta_\mathrm{rec}\cdot d_\mathrm{wire}\cdot \frac{E_\mathrm{rec}}{2},
    \end{equation}
where $j_{\rm{HABS}}$ is the shape function of the beam evaluated at the angular position $\theta(i)$ of the wire  element, $\Phi_{\rm{at}}$ is the total flow of atoms in the beam (atoms/second) and $E_\mathrm{rec}$ is the energy released when two hydrogen atoms recombine into a molecule. $\gamma_\mathrm{rec}$ is the likelihood of an atom which  hits the wire to recombine on the wire, $ \beta_\mathrm{rec}$ is the fraction of the recombination energy  which is transferred to the wire in case of a molecule is produced. These last two parameters are poorly constrained but some rough values are available in the literature\cite{Melin1970}. For the  example shown in Figure \ref{fig:Heat_flow} $\gamma_\mathrm{rec}$ and $ \beta_\mathrm{rec}$ are set to 1, since it will later simply become part of the fit parameter $A$. Choosing the factor 1 minimizes the atom flux required to explain the recombination  signal that is measured, allowing for the determination of a lower limit on the atoms actually produced. This is desirable because producing a large number of atoms is a requirement for the goals of the Project 8 collaboration.
    
- $f_\mathrm{bb}$: Blackbody radiation from the HABS.
    
    \begin{equation}
        f_\mathrm{bb} = \varepsilon \cdot \sigma \cdot T_\mathrm{HABS}^4\cdot A_\mathrm{HABS}\frac {l\cdot d_\mathrm{wire}}{4\pi r_\mathrm{HABS}^2}\frac 1 l,
    \end{equation}
where $\sigma$ is the Stephan-Boltzmann-constant, $\epsilon$ is the emissivity and $A_\mathrm{HABS}$ is the area of the source visible to the wire, $r_\mathrm{HABS}$ is the distance to the HABS from the wire. The fraction of $\dfrac{A_\mathrm{HABS}}{4\pi r_\mathrm{HABS}^2}$ yields the portion of the field of view as seen from the wire which is glowing hot at source temperatures.

\section{Wire sensitivity}
\label{sec:wire_sensitivity}

Using the simulation described in Section \ref{sec:Signal components}, a test heating power can be applied to an arbitrary point of a simulated wire in thermal equilibrium. By calculating the total resistance of the wire with and without the test point source, the change in resistance due to the test source is determined. The position at  which the test source is applied is then shifted along the wire, until a sample is generated at every segment of the simulated wire. This allows for determining the relative effect of a heating power based on the location at which it is applied to the wire. The absolute size of the simulated $\Delta R$ signal is deemed unreliable at worse than 10\% accuracy, however the relative effect size should be much better than that.

In practice, a simulated 1 $\rm {\mu W}$ point source is applied along the wire and $\Delta R$ induced in the wire resistance is recorded. The values of $\Delta R$ are then normalized by the average value $\Delta R_{avg}$ yielding the relative wire sensitivity

\begin{equation}
\label{eq:eta_wire}
\eta_{\rm{wire}}(x_i) = \frac{\Delta R (x_i)}{\Delta R_{avg}}
\end{equation}
with $x_i$ the position of the wire of segment $i$ at which the simulated point heat source is applied.

\begin{figure}[H]
\centering
	\includegraphics[width=1\linewidth]{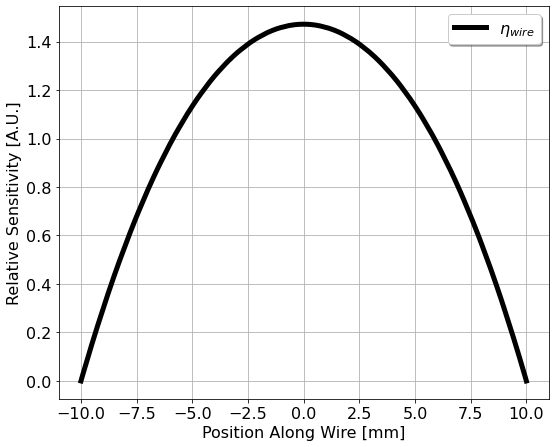}
    	\caption{Normalized local wire sensitivity $\eta_{wire}$. A larger $\eta$ indicates a larger resistance change for a  given  change in input power. Sensitivity is highest in the center of the wire with steepening decline to the ends of the wire.
     }
	\label{fig:eta_wire}
\end{figure}

Sensitivity is highest in the center of the wire with steepening decline to the ends of the wire. 
The change in local wire sensitivity is caused by the fact that the temperature profile is curved leading to places where (primarily) thermal conduction is more or less effective at carrying extra heat out of the wire. Figure \ref{fig:eta_wire} shows the value of the relative wire sensitivity along the wire. 

When using the wire this means that the part of the beam hitting  the center of  the wire has a disproportionate effect.
This is accounted for by  interpolating over the $\eta_{\rm{wire}}(x_i)$ of each segment and including the resulting function $\eta_{\rm{wire}}(x)$ in the signal model given  in  Eq. \ref{eq:P_hit_fit}. 

Efforts to directly measure the wire sensitivity using a laser as a point source of heat are ongoing, but not included in this publication.

\section{Correcting calibration for temperature distribution and resulting wire sensitivity}
\label{sec:correcting calibration}

The calibration method initially assumes that the wire is heated equally in all places by the measurement current. This is incorrect, but a convenient approximation, because  only the total resistance of the wire is measurable and therefore only an  approximation for its average temperature can be derived.

A better approximation can be obtained,  by simulating the wire  behavior.

\begin{figure}[H]
\centering
	\includegraphics[width=1\linewidth]{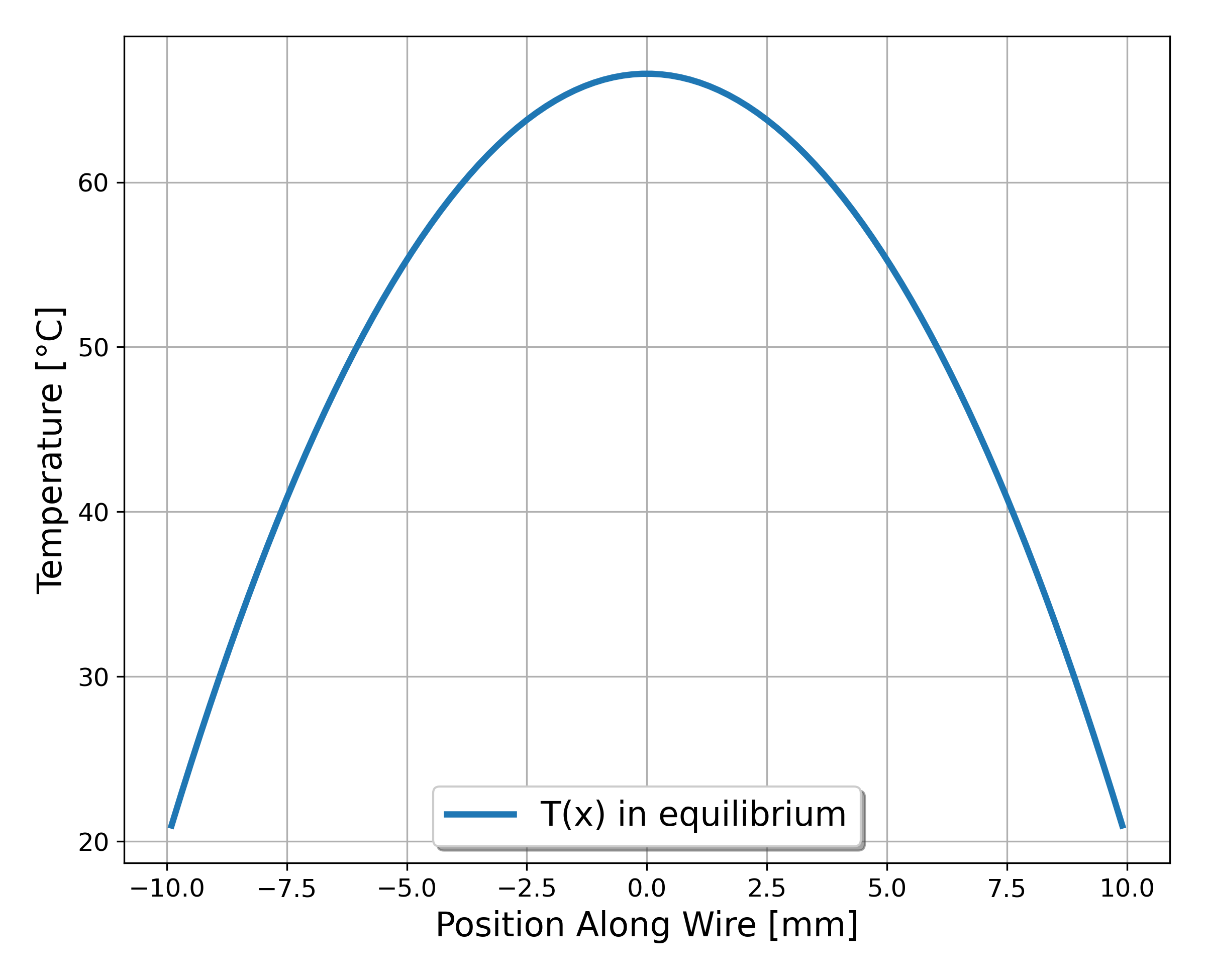}
    	\caption{Simulated temperature profile along the wire at 1mA measurement current. The maximum of the wire temperature is reached in the center of the wire at just under 70 °C. There is a roughly 45 °C temperature difference to the edges of the wire.}
	\label{fig:sim_temp_profile}
\end{figure}

The finite elements simulation results in a prediction for the wire temperature profile in response to the heating due to the measurement current, shown in Figure \ref{fig:sim_temp_profile}.

Of course the real calibration measurement already includes these effects. What remains for us is to correct the assumption, that $k$ as calculated above is the average conversion factor for $\Delta R$ to power. As calculated, it is actually already the conversion factor weighted for the unequally distributed heating power. To get back to the  average conversion factor for a flat input power the effect of the additional central heating needs to be removed.

The difference in global calibration factor based on the real input power distribution and a flat input power can be interpreted as a correction factor $c$, the quotient of the integrals over the power distribution weighted by $\eta_{wire}$ along  the wire

\begin{equation}
\label{eq:c}
 c = \frac{\int_{wire} I^2 R_{dens}(T(x)) \cdot \eta_{wire}(x) dx}
    {\int_{wire} I^2 \frac{R_{tot}}{l_{wire}} \cdot \eta_{wire}(x) dx },
\end{equation}
which simplifies to 
\begin{equation}
\label{eq:c2}
c = \frac{\int_{wire} R_{dens}(T(x)) \cdot \eta_{wire}(x) dx}
    {\frac{R_{tot}}{l_{wire}} \int_{wire} \eta_{wire}(x) dx }.
\end{equation}
and finally
\begin{equation}
\label{eq:c3}
c = \frac{\int_{wire} (1 + a\cdot \Delta T(x)) \cdot \eta_{wire}(x) dx}
   { \int_{wire} \eta_{wire}(x) dx },
\end{equation}
where $a$ is the temperature coefficient of resistivity and $\Delta T = T(x) - T_{avg}$. $T_{avg}$ being the average wire temperature that would result in it having  a total resistance of $R_{tot}$.
The correction resulting from a simulation at $1\,\rm{mA}$ is $c = 1.0228$, meaning that the calculated $k_{eff} = c \cdot k$ is too high  by  this factor compared to the equally distributed heating power assumption. The accuracy of this correction depends on the accuracy of the simulation, but since it is a fairly small correction, even large inaccuracies in the simulation on the order of 10\% would only translate into part per thousand level changes to the calibration factor $k$.

\section{Axes definition}
\label{sec:Axes Definition}
\begin{figure}[htbp]
\centering
	\includegraphics[width=1\linewidth]{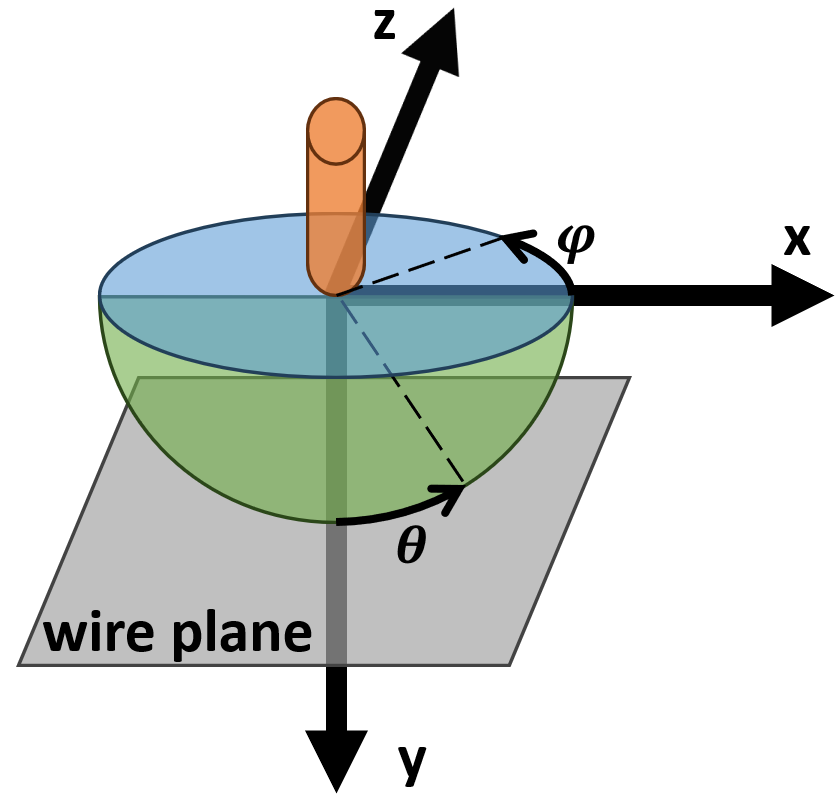}
    	\caption{Half-sphere axes definitions. The HABS capillary is indicated by the orange tube. The wire is oriented parallel to the x-axis in the wire plane.  It can be moved in the z-direction by a translation stage. The distance to the capillary in the y-direction is fixed. 
     }
	\label{fig:axes_schematic}
\end{figure}

\section{Effective length over flow}
\label{sec:Appendix_leff_over_flow}

This section contains the plots of extracted $P_{\rm rec}$ for 0.2 sccm and 0.05 sccm along with fits. The method is equivalent to the three-temperature-point method demonstrated for a flow of 1 sccm in Section \ref{sec:Application to Data}. The resulting values for $\leff$ are included in Figure \ref{fig:leff_over_flow}.

\begin{figure}[htbp]
    \centering
    \includegraphics[width=1\linewidth]{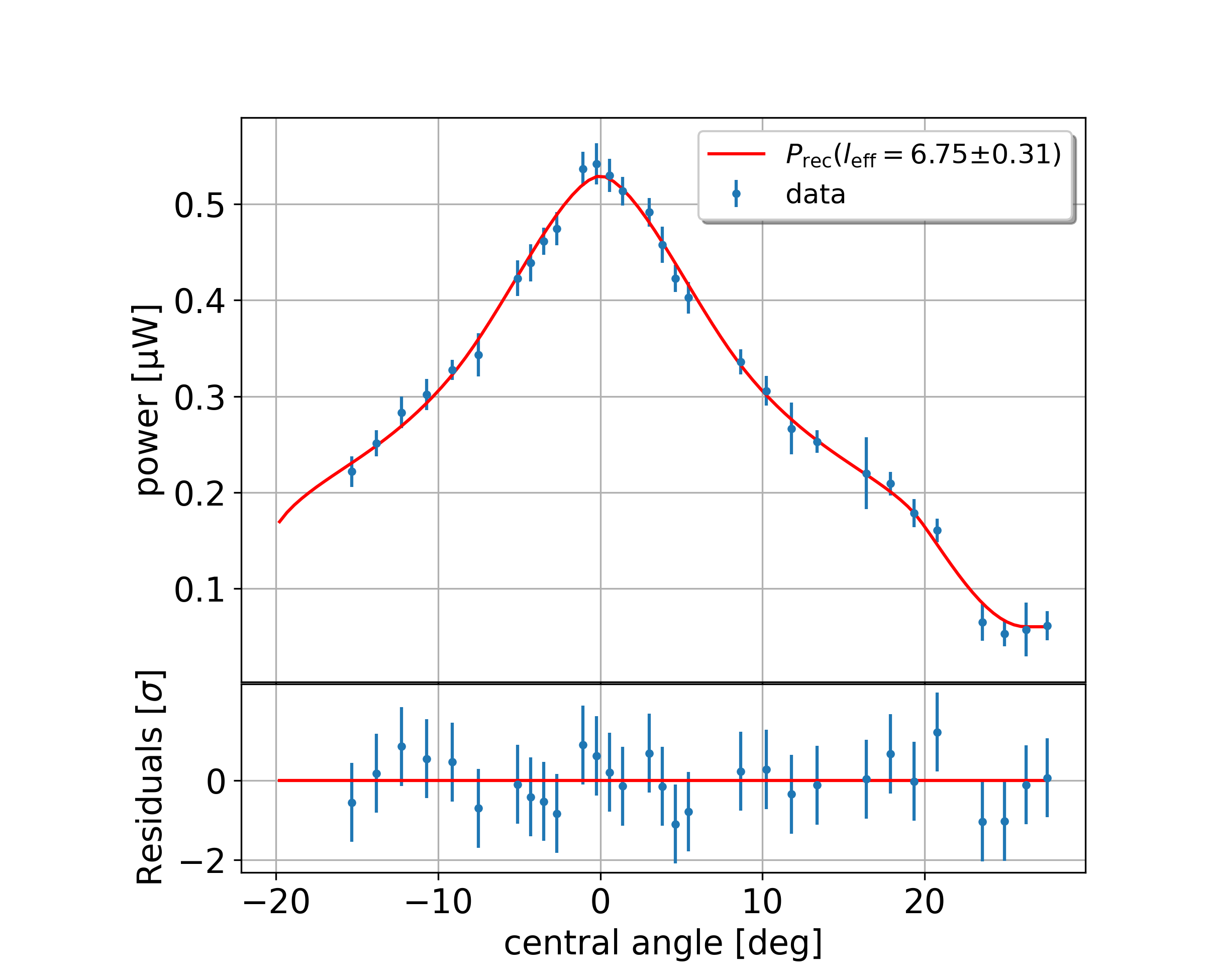}
    \caption{Extracted heating power attributed to recombination ($P_{\rm{rec}}$) for a flow of $0.2 \, \rm{sccm}$ at $2211 \,\rm{K}$ source temperature, shown over the range of positions the wire was placed at. The optimized beam shape model is plotted alongside. $P_{H_2} $(Eq. \eqref{eq:P_H2_fit}) was determined using z-scans 1277~K and 298~K.  }
    \label{fig:/fit_02sccm}
\end{figure}

\begin{figure}[htbp]
    \centering
    \includegraphics[width=1\linewidth]{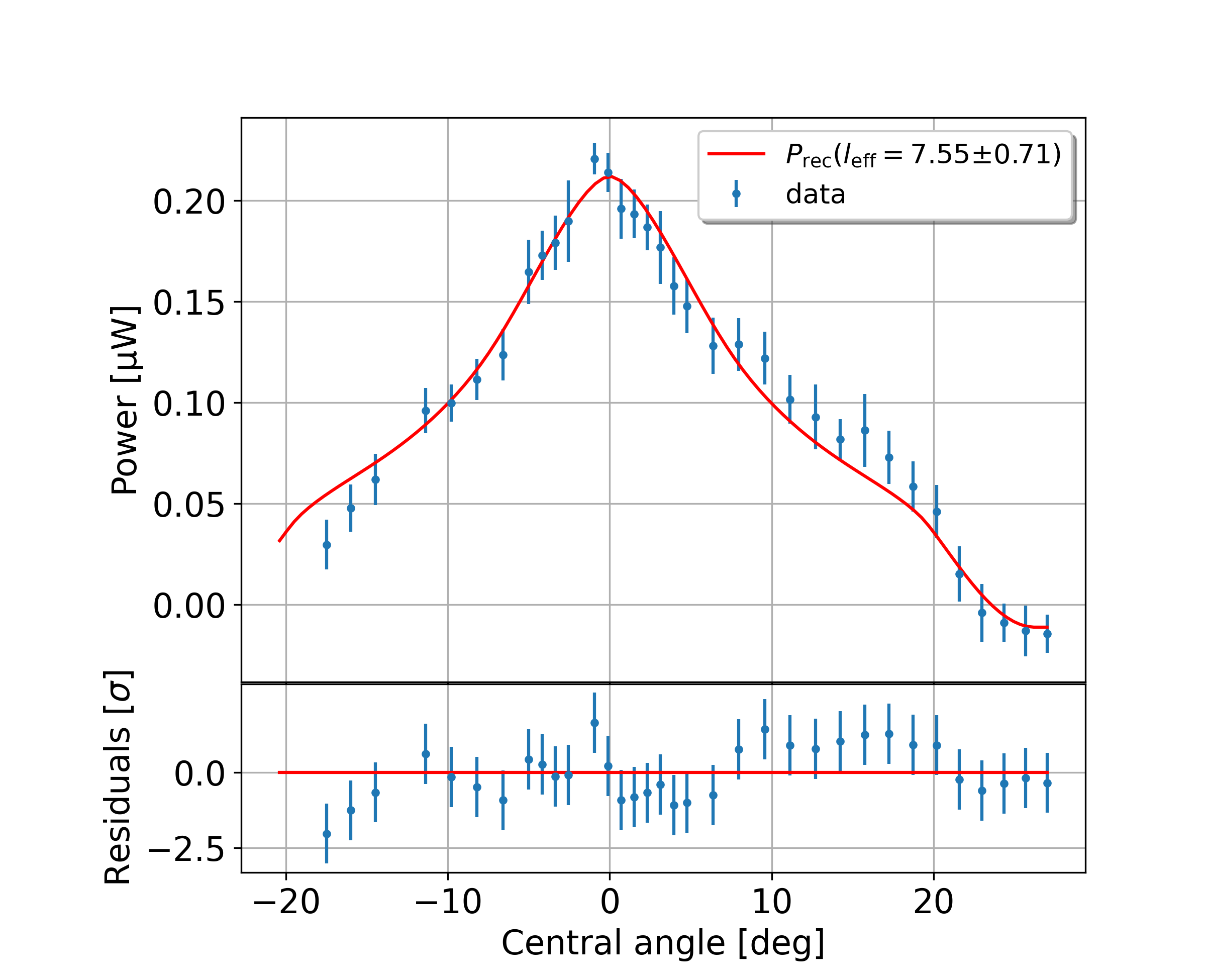}
    \caption{Extracted heating power attributed to recombination ($P_{\rm{rec}}$) for a flow of $0.05 \, \rm{sccm}$ at $2211 \,\rm{K}$ source temperature, shown over the range of positions the wire was placed at. The optimized beam shape model is plotted alongside. $P_{H_2} $(Eq. \eqref{eq:P_H2_fit}) was determined using z-scans 1277~K and 298~K. }
    \label{fig:fit_005sccm}
\end{figure}

\section{Systematic errors from temperature selection}
\FloatBarrier
\label{sec:systematics_T}

\begin{figure}[htbp]
    \centering
    \includegraphics[width=1\linewidth]{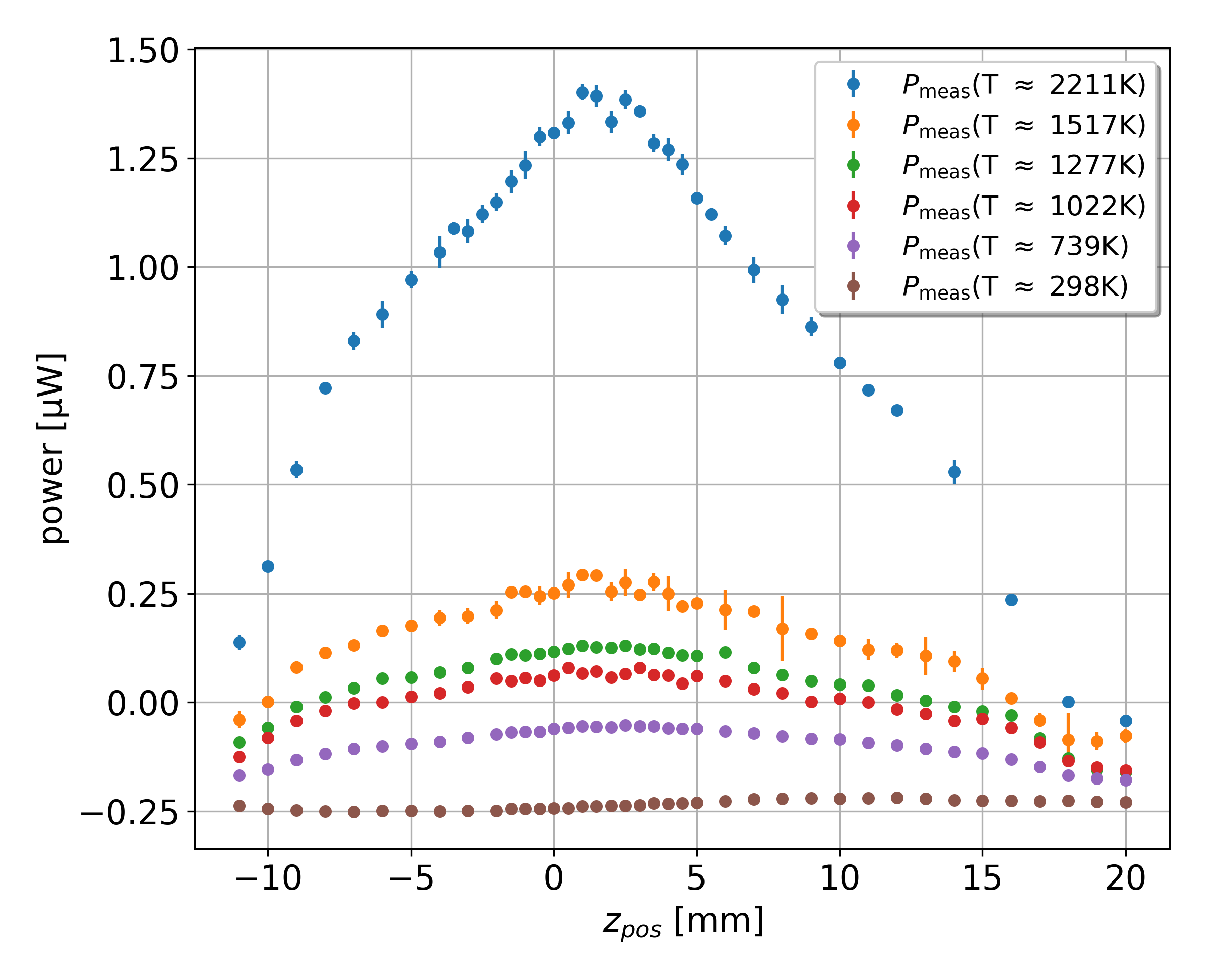}
    \caption{Power detected on the wire after primary extraction ($P_{\rm{meas}}$) at $1 \, \rm{sccm}$ for an extended range of source temperatures. All datasets below 1600~K are below the dissociation threshold and measure a pure $\rm H_2$ beam.  These data sets are then used in the secondary extraction procedure, to yield the effect of recombining atoms on the surface. Negative detected powers indicate a net cooling of the wire, which is identified with the convection cooling due to gas pressure in the detector chamber ($f_\mathrm{bkgd\_gas}$). }
    \label{fig:6T_1sccm}
\end{figure}

For the 1 sccm flow setting we took additional z-scans at additional intermediate temperatures below the dissociation threshold. Additional scans were not taken for every flow setting due to time constraints in the availability of the setup. Each of the z-scans with 38 positions requires about 80 hours to measure with the procedure presented in Figure \ref{fig:1sccm_signal_extraction}, such that the 6 datasets in Figure \ref{fig:6T_1sccm} represent over 20 days of runtime alone. The three-temperature-point method presented in this paper is a deliberate minimum of runtime, to make it possible to search in other parts of the parameter space, such as flow.

\begin{figure}
    \centering
    \includegraphics[width=1\linewidth]{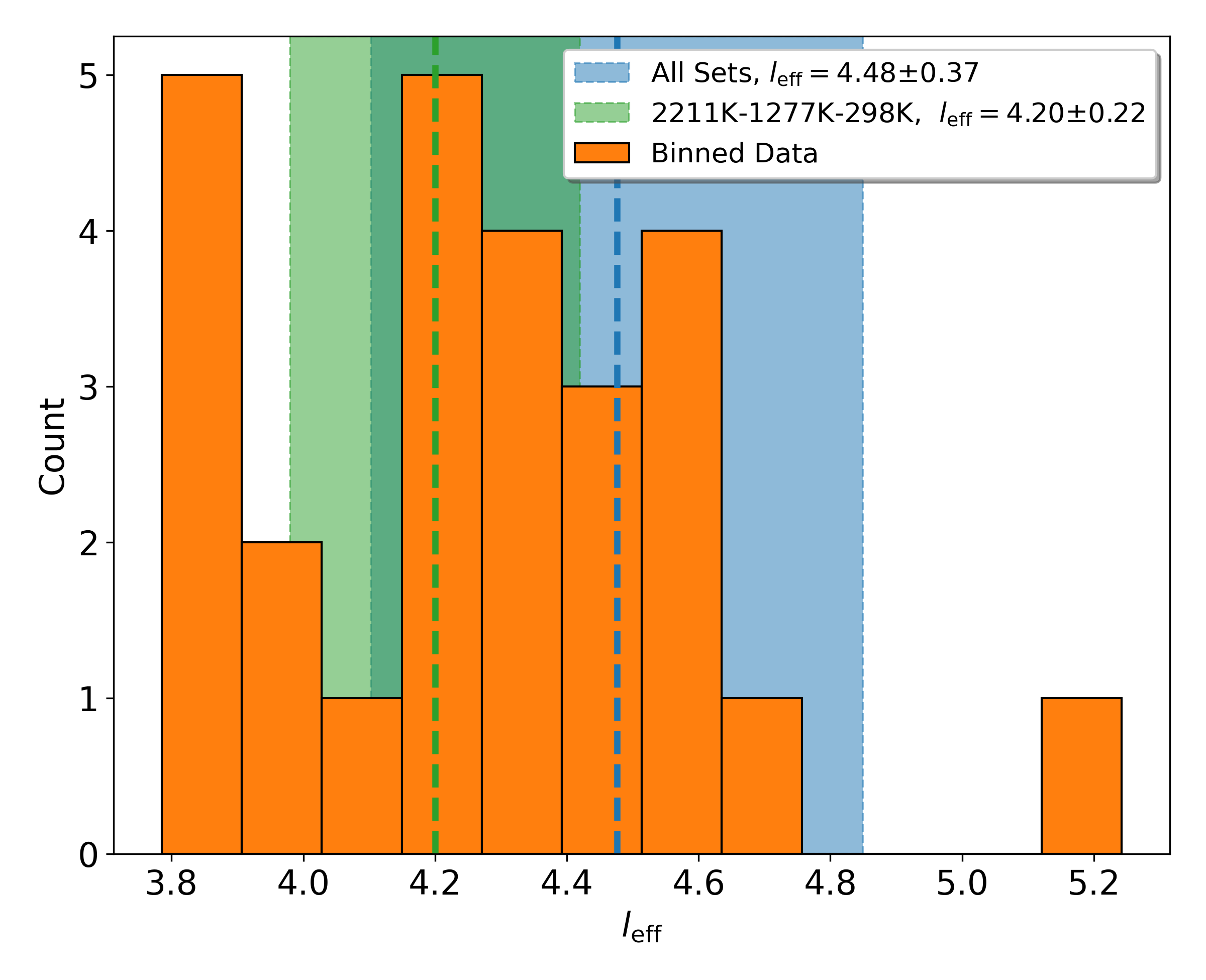}
    \caption{Histogram of fit parameter $\leff$ when using all possible combinations of the sub-dissociation datasets to fit $P_{H_2}$ (Eq. \eqref{eq:P_H2_fit}). The results of the fits using all available sets at once (Figure \ref{fig:fit_1sccm_all}) and the three-temperature-method as presented in Section \ref{sec:Application to Data} are displayed with error bands.
    }
    \label{fig:Histogram}
\end{figure}

\begin{figure}
    \centering
    \includegraphics[width=1\linewidth]{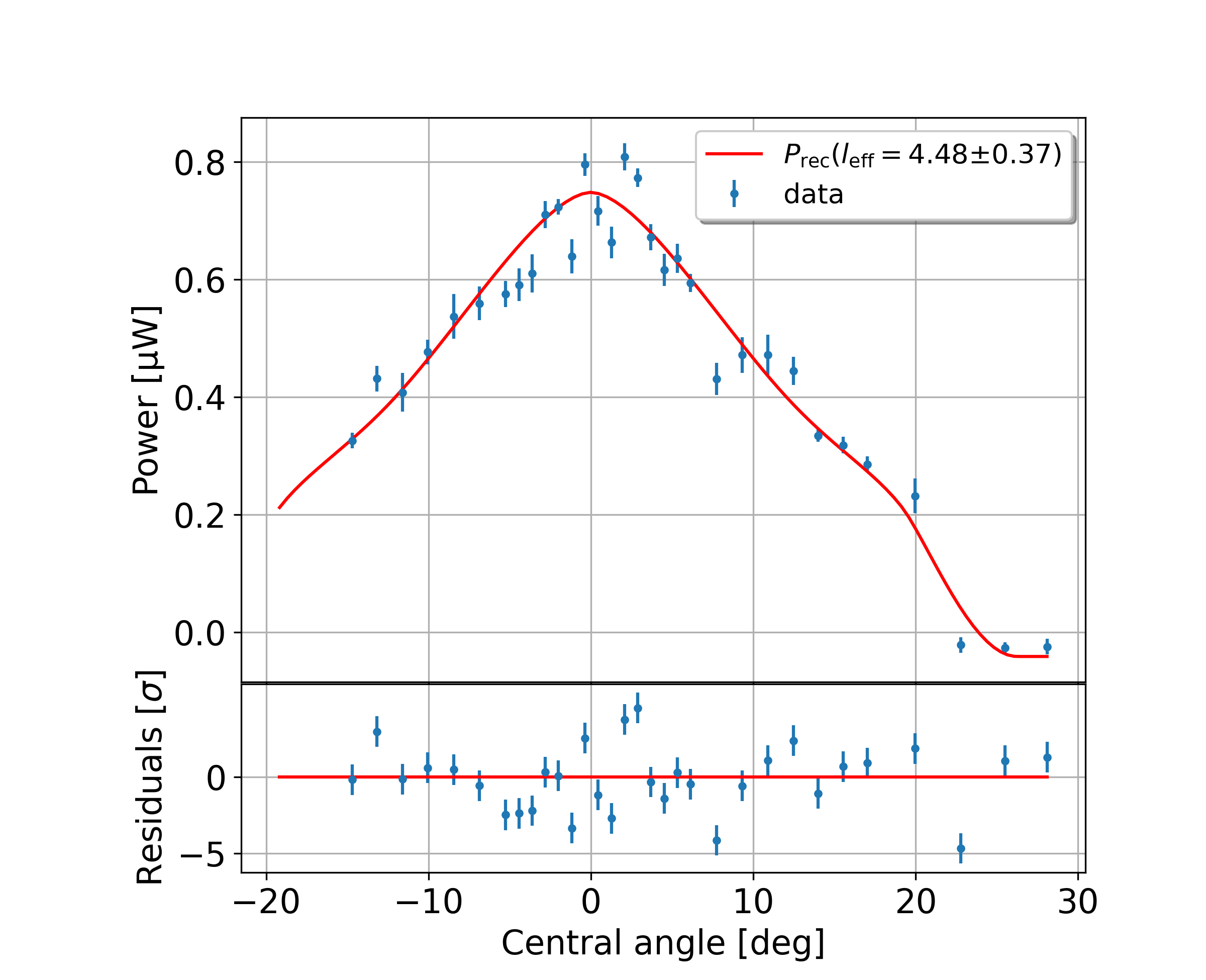}
    \caption{Extracted heating power attributed to recombination ($P_{\rm{rec}}$) for a flow of $1 \, \rm{sccm}$ at $2211 \,\rm{K}$ source temperature, shown over the range of positions the wire was placed at. The optimized beam shape model is plotted alongside. $P_{H_2} $(Eq. \eqref{eq:P_H2_fit}) was determined using all available z-scans below the dissociation threshold: 1518~K, 1277~K, 1022~K, 739~K and 298~K. }
    \label{fig:fit_1sccm_all}
\end{figure}

We use the additional z-scans to gauge possible systematic deviation in fit parameters due to the arbitrary choice of temperatures below the dissociation threshold we used to fit Eq. \ref{eq:P_H2_fit}. The secondary extraction procedure (Section \ref{sec:secondary_extraction}) is performed using all possible combinations (containing between 2-5) of the non-dissociating temperature setpoints (the set {1518~K, 1277~K, 1022~K, 739~K, 298~K}). The sole dissociating temperature setpoint 2211~K must of course always be included as $T_{high}$. The resulting extracted powers are then fit with Eq. \ref{eq:P_hit_fit_long}. The resulting $\leff$ fit parameter of all 26 combinations is shown as a histogram in Figure \ref{fig:Histogram}. The result of the three-point-method as used in this paper is shown in green along with its error band. Similarly the result of the extraction and fit based on all 6 available datasets at once is shown in blue. The extracted recombination power and its fit are shown in Figure \ref{fig:fit_1sccm_all}. The value of $\leff = 4.48 \pm 0.37$ is statistically compatible with the three-point-method result previously discussed. 


This histogram is an imperfect proxy for systematic error due to temperature setpoint selection. The entries are clearly not entirely statistically independent as only 6 independent datasets exist. It is therefore unsurprising that the histogram does not approximate a normal distribution well. Nevertheless it illustrates that three-point-method using the 2211~K , 1277~K and 298~K datasets presented in this work is a reasonable expedient. For the systematic error bar displayed in Figure \ref{fig:leff_over_flow} we use the span of all results for $\leff$ after excluding the outlier at 5.24. This outlier is produced when extrapolating from the 1277~K and 1022~K datasets. It is perhaps unsurprising, that using datasets that are adjacent in temperature are more prone to large variation due to their shorter temperature baseline. The resulting span range for is 3.79 to 4.67, which leaves the three-point-method result of $\leff = 4.20 \pm 0.22$ well centered within. This span is plotted in orange as a systematic error bar in Figure \ref{fig:fit_1sccm_2250K_penumbra}.

\acknowledgments
This material is based upon work supported by the following sources: the Cluster of Excellence "Precision Physics, Fundamental Interactions, and Structure of Matter" (PRISMA+ EXC 2118/1) funded by the German Research Foundation (DFG) within the German Excellence Strategy (Project ID 39083149); the U.S. Department of Energy Office of Science, Office of Nuclear Physics, under Award No.~DE-SC0020433 to Case Western Reserve University (CWRU), under Award No.~DE-SC0011091 to the Massachusetts Institute of Technology (MIT), under Field Work Proposal Number 73006 at the Pacific Northwest National Laboratory (PNNL), a multiprogram national laboratory operated by Battelle for the U.S. Department of Energy under Contract No.~DE-AC05-76RL01830, under Early Career Award No.~DE-SC0019088 to Pennsylvania State University, under Award No.~DE-SC0024434 to the University of Texas at Arlington, under Award No.~DE-FG02-97ER41020 to the University of Washington, and under Award No.~DE-SC0012654 to Yale University; the National Science Foundation under Grant No.~PHY-2209530 to Indiana University, and under Grant No.~PHY-2110569 to MIT; the Karlsruhe Institute of Technology (KIT) Center Elementary Particle and Astroparticle Physics (KCETA); Laboratory Directed Research and Development (LDRD) 18-ERD-028 and 20-LW-056 at Lawrence Livermore National Laboratory (LLNL), prepared by LLNL under Contract DE-AC52-07NA27344, LLNL-JRNL-868753; the LDRD Program at PNNL; and Yale University.

\section*{Author Declarations}
\subsection*{Conflicts of Interest}
The authors have no conflicts to disclose.
\subsection*{Author Contributions (CRediT)}
\noindent\textbf{C. Matthé}: Conceptualization, Methodology (lead), Data curation, Formal analysis (lead), Investigation (lead), Software, Validation (lead), Visualization (lead), Writing – original draft (lead), Writing – review \& editing (lead). \\
\textbf{S. Böser}: Conceptualization, Funding acquisition, Project administration, Resources, Supervision. \\
\textbf{M. Fertl}: Funding acquisition, Resources. \\
\textbf{L. A. Thorne}, \textbf{D. Fenner}, \textbf{M. B. Hüneborn}, \textbf{M. Astashov}, \textbf{P. Kern}, \textbf{A.~Lindman}, \textbf{B. Mucogllava}: Investigation (supporting). \\
\textbf{S. Enomoto}, \textbf{N.S. Oblath}, \textbf{W. Pettus}, \textbf{M. Astashov}, \textbf{A. Lindman}, \textbf{L. A. Thorne}: Software, Data curation. \\
\textbf{B. J. P. Jones}, \textbf{R. G. H. Robertson}, \textbf{B. Mucogllava}, \textbf{M.~Wynne}, \textbf{L. A. Thorne}, \textbf{T. E. Weiss}, \textbf{W. Pettus}, \textbf{M.~Oueslati}, \textbf{J.~A.~Formaggio}, \textbf{A. Lindman}, \textbf{M.~J.~Brandsema}, \textbf{A.~L.~Reine}: Review \& Editing. \\
\textbf{K. Kazkaz}: Review \& Editing, Validation (supporting). \\
\textbf{All other authors}: Review \& Editing (supporting).


\section*{Data Availability Statement}
The data that support the findings of this study are available from the corresponding authors upon reasonable request.




\nocite{*}
\bibliography{biblio.bib}

\end{document}